\documentclass[twoside,twocolumn,english,prc,amsmath,a4paper,myheadings]{revtex4}
\usepackage{bookman}
\usepackage[T1]{fontenc}
\usepackage[latin9]{inputenc}
\usepackage{geometry}
\usepackage{verbatim}
\usepackage{calc}
\usepackage{textcomp}
\usepackage{amsmath}
\usepackage{amssymb}
\usepackage{graphicx}
\PassOptionsToPackage{normalem}{ulem}
\usepackage{ulem}

\makeatletter

\providecommand{\tabularnewline}{\\}

\@ifundefined{textcolor}{}
{%
 \definecolor{BLACK}{gray}{0}
 \definecolor{WHITE}{gray}{1}
 \definecolor{RED}{rgb}{1,0,0}
 \definecolor{GREEN}{rgb}{0,1,0}
 \definecolor{BLUE}{rgb}{0,0,1}
 \definecolor{CYAN}{cmyk}{1,0,0,0}
 \definecolor{MAGENTA}{cmyk}{0,1,0,0}
 \definecolor{YELLOW}{cmyk}{0,0,1,0}
}

\def\@oddhead{\rightmark \hfill Charm production in high muliplicity pp events  \hfill \thepage}
\def\@evenhead{\thepage \hfill K. Werner et al.\hfill}
\def\fnum@table{\tablename~{\bf\thetable}}
\def\fnum@figure{\figurename~{\bf\thefigure}}
\def\tablename{\footnotesize{\bf Table}}
\def\figurename{\footnotesize{\bf Figure}}

\usepackage{dcolumn}

\def\citet{\cite}

\AtBeginDocument{

}

\makeatother

\usepackage{babel}
\begin{document}

\title{\noindent Charm production in high multiplicity pp events}

\author{{\normalsize{}K.$\,$Werner$^{(a)}$, B. Guiot$^{(a,b)}$, Iu.$\,$Karpenko$^{(c,d,e)}$,
T.$\,$Pierog$^{(f)}$ , G. Sophys$^{(a)}$\medskip{}
}}

\address{$^{(a)}$ SUBATECH, University of Nantes -- IN2P3/CNRS-- EMN, Nantes,
France}

\address{$^{(b)}$ Universidad Técnica Federico Santa María, Valparaiso ,
Chile}

\address{$^{(c)}$ FIAS, Johann Wolfgang Goethe Universitaet, Frankfurt am
Main, Germany}

\address{$^{(d)}$ Bogolyubov Institute for Theoretical Physics, Kiev 143,
03680, Ukraine}

\address{$^{(e)}$ INFN - Sezione di Firenze, Via G. Sansone\,1, I-50019
Sesto Fiorentino (Firenze), Italy}

\address{$^{(f)}$Karlsruhe Inst. of Technology, KIT, Campus North, Inst.
f. Kernphysik, Germany}
\begin{abstract}
Recent experimental studies of the multiplicity dependence of heavy
quark (HQ) production in proton-proton collisions at 7 TeV showed
a strong non-linear increase of the HQ multiplicity as a function
of the charged particle multiplicity. We try to understand this behavior
using the EPOS3 approach. Two issues play an important role: multiple
scattering, in particular its impact on multiplicity fluctuations,
and the collective hydrodynamic expansion. The results are very robust
with respect to many details of the modeling, which means that these
data contain valuable information about very basic features of the
reaction mechanism in proton-proton collisions.
\end{abstract}
\maketitle

\section{Introduction}

Although being very rare, high multiplicity events in high energy
proton-proton collisions gained much attention in recent years, after
discovering many features which look very much like the collective
effects known from heavy ion collisions, in particular azimuthal anisotropies
or the mass dependence of particle spectra. One usually studies effects
as a function of the event activity, the latter one being typically
the charged particle multiplicity, which takes the role of the ``centrality''
in heavy ion studies. 

\begin{figure}[b]
\noindent \begin{centering}
\includegraphics[angle=270,scale=0.25]{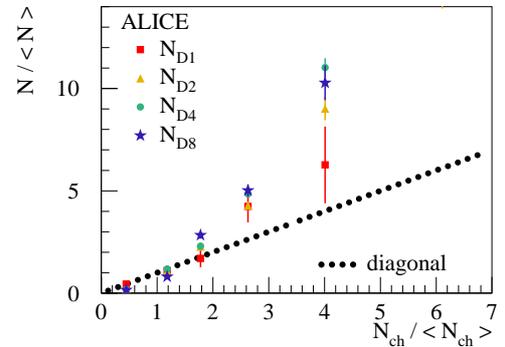}
\par\end{centering}

\protect\caption{(Color online) $D$ meson multiplicities versus the charged particle
multiplicity, both divided by the corresponding minimum bias mean
values. The different symbols and the notations $N_{D1}$, $N_{D2}$,
$N_{D4}$, $N_{D8}$ refer to different $p_{t}$ ranges: 1-2, 2-4,
4-8, 8-12 (in GeV), $N_{\mathrm{ch}}$ refers to the charged particle
multiplicity. \label{fig:NDvsNch_data}}

\end{figure}

Recently, several experimental groups investigated the dependence
of heavy quark production on the event activity, both for open and
hidden charm or bottom. We will focus here on $D$ meson production,
where the term ``$D$ meson multiplicity'' refers in the following
to the average multiplicity of $D^{+}$, $D^{0}$ and $D^{*+}$. The
ALICE collaboration found a quite unexpected result \cite{alice1}
as shown in fig. \ref{fig:NDvsNch_data}: When plotting the $D$ meson
multiplicity versus the charged particle multiplicity, both divided
by the corresponding minimum bias mean values, one obtains a dependence
which is very significantly more than linear, the latter one indicated
by the dotted line (referred to as ``diagonal'', meaning $N_{D}/\left\langle N_{D}\right\rangle =N_{\mathrm{ch}}/\left\langle N_{\mathrm{ch}}\right\rangle $).
The effect seems to be bigger for larger $p_{t}$.
Both $D$ meson and charged particle multiplicity refer to central
rapidities. 

Trying to understand their data, the authors of \cite{alice1} analyze
simulations using PYTHIA 8.157. Taking only $D$ mesons from the hard
process, the dependence on the charged particle multiplicity $N_{\mathrm{ch}}$
is even much less than linear, taking into account all contributions
(in particular multiple scattering), the results are close to the
diagonal, slightly below for small $p_{t}$ and slightly above for
high $p_{t}$. So multiple scattering seems to help, it goes into
the right direction, but the simulations are still far from the data.

In \cite{alice1}, they also refer to a percolation approach \cite{perco},
which assumes that high-energy hadronic collisions are driven by the
exchange of color sources between the projectile and target in the
collision. These color sources have a finite spatial extension and
can interact. In a high-density environment, the coherence among the
sources leads to a reduction of their effective number. The source
transverse mass determines its transverse size, and allows to distinguish
between soft and hard sources. As a consequence, at high densities
the total charged-particle multiplicity, which originates from soft
sources, is reduced. In contrast, hard particle production is less
affected due to the smaller transverse size of hard sources. The percolation
model predicts a faster-than-linear increase of heavy flavor relative
production with the relative charged-particle multiplicity. However,
there is no discussion the the $p_{t}$ dependence.

In \cite{alice1}, they also show results from the multiple scattering
approach EPOS3 \cite{epos3}. It is a universal model, using the same
scheme for pp, pA, and AA collisions, in particular a hydrodynamic
expansion stage. It is shown in the paper, that EPOS without hydro
get a slightly more than linear increase, with a strong enhancement
when considering the hydro expansion. This is understood due to the
fact that in case of hydrodynamical expansion, the multiplicity is
considerably reduced compared to the case without.

Considering all these results, on may conclude (in a qualitative fashion):
It seems that multiple scattering is crucial, without one would not
even get close to a linear behavior. A second aspect seems to be important:
A reduction of the charged particle multiplicity (rather than an increase
of the $D$ meson multiplicity) due to collective effects.

It is clear that the experimental observations are very interesting,
and provide valuable insight into the very nature of the reaction
mechanism in $pp$ scattering, in particular in case of high event
activity. So we try in this paper to provide a detailed analysis of
the phenomenon in the EPOS3 framework. Two aspects provide the key
to the understanding: Multiple scattering and collectivity.

\section{A short EPOS3 summary}

\begin{figure}
\fbox{\begin{minipage}[t]{1\columnwidth}%
\noindent {\large{}~}{\large \par}

\noindent {\large{}\hspace*{-1cm}}%
\begin{minipage}[c]{0.35\columnwidth}%
\noindent \begin{center}
{\large{}
\[
\sigma^{\mathrm{tot}}\negthickspace=\negthickspace\sum_{\mathrm{cut\,P}}\int\negthickspace\sum_{\mathrm{uncut\,P}}\int
\]
}
\par\end{center}{\large \par}%
\end{minipage}%
\begin{minipage}[c]{0.5\columnwidth}%
\noindent \begin{center}
\includegraphics[scale=0.22]{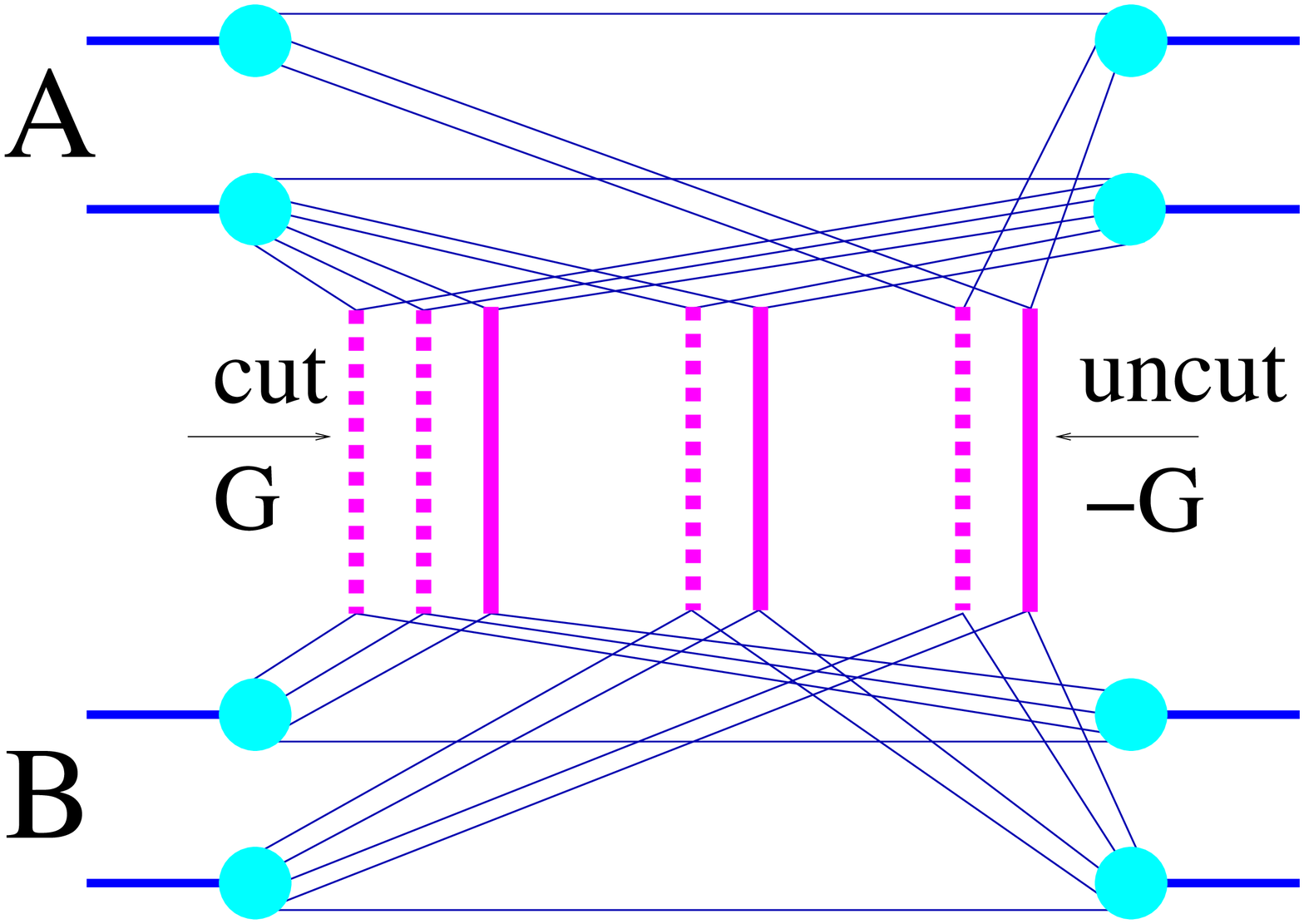}
\par\end{center}%
\end{minipage}\vspace{-0.2cm}

\noindent \begin{center}
\begin{minipage}[t]{0.95\columnwidth}%
\[
\,\,\,\,\qquad\quad\qquad\underbrace{\qquad\qquad\qquad\quad\qquad\qquad\qquad\qquad\qquad\qquad}
\]
\vspace{-0.9cm}
\end{minipage}
\par\end{center}

\begin{minipage}[t]{0.95\columnwidth}%
{\large{}
\[
\qquad\qquad\qquad d\sigma_{\mathrm{exclusive}}
\]
}~%
\end{minipage}

~%
\end{minipage}}

\protect\caption{(Color online) PBGRT formalism: The total cross section expressed
in terms of cut (dashed lines) and uncut (solid lines) Pomerons, for
nucleus-nucleus, proton-nucleus, and proton-proton collisions. Partial
summations allow to obtain exclusive cross sections, the mathematical
formulas can be found in \cite{hajo}, or in a somewhat simplified
form in \cite{epos3}. \label{muscatt}}
\end{figure}

\begin{figure*}
\noindent \includegraphics[scale=0.25]{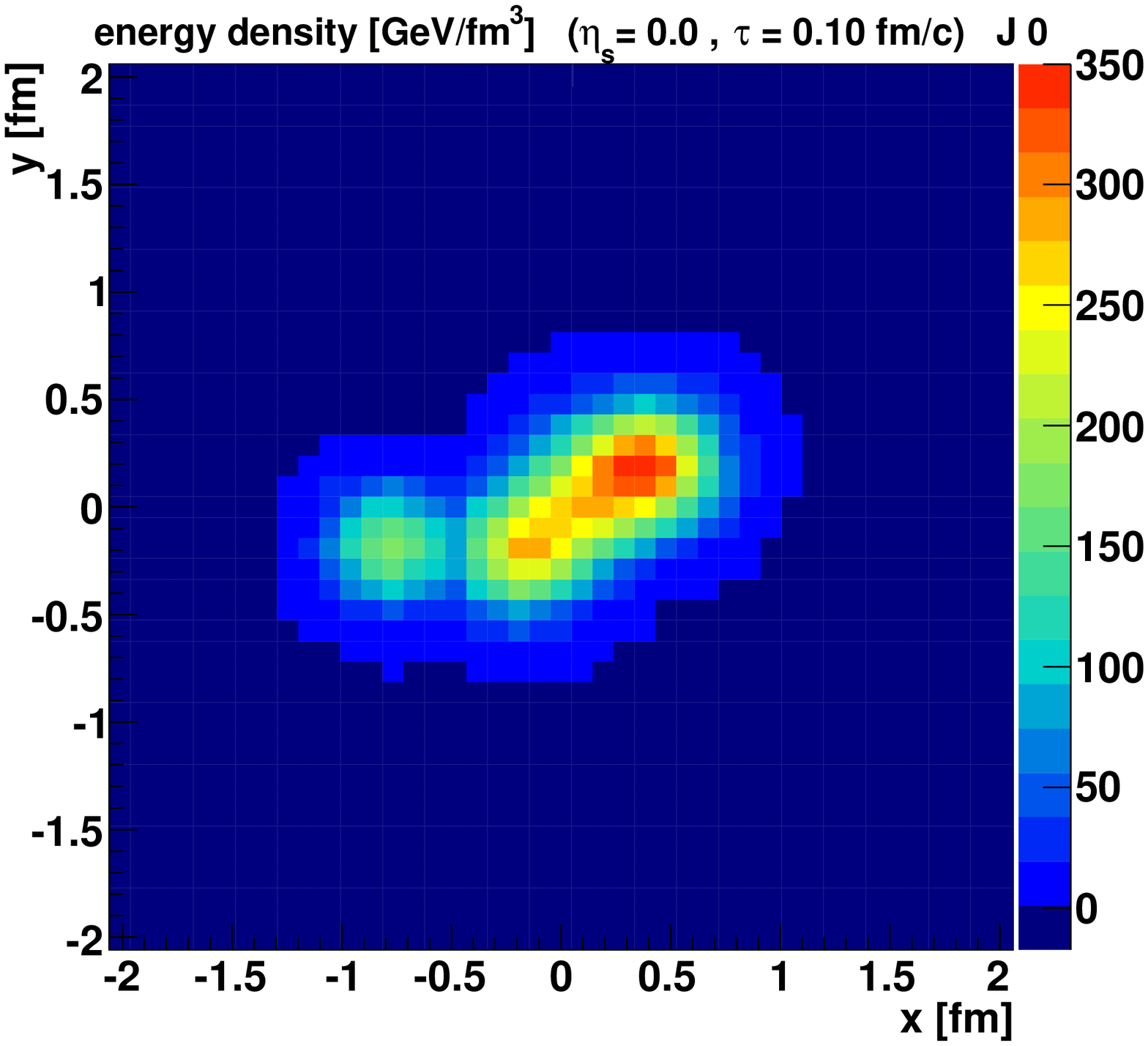}\includegraphics[scale=0.25]{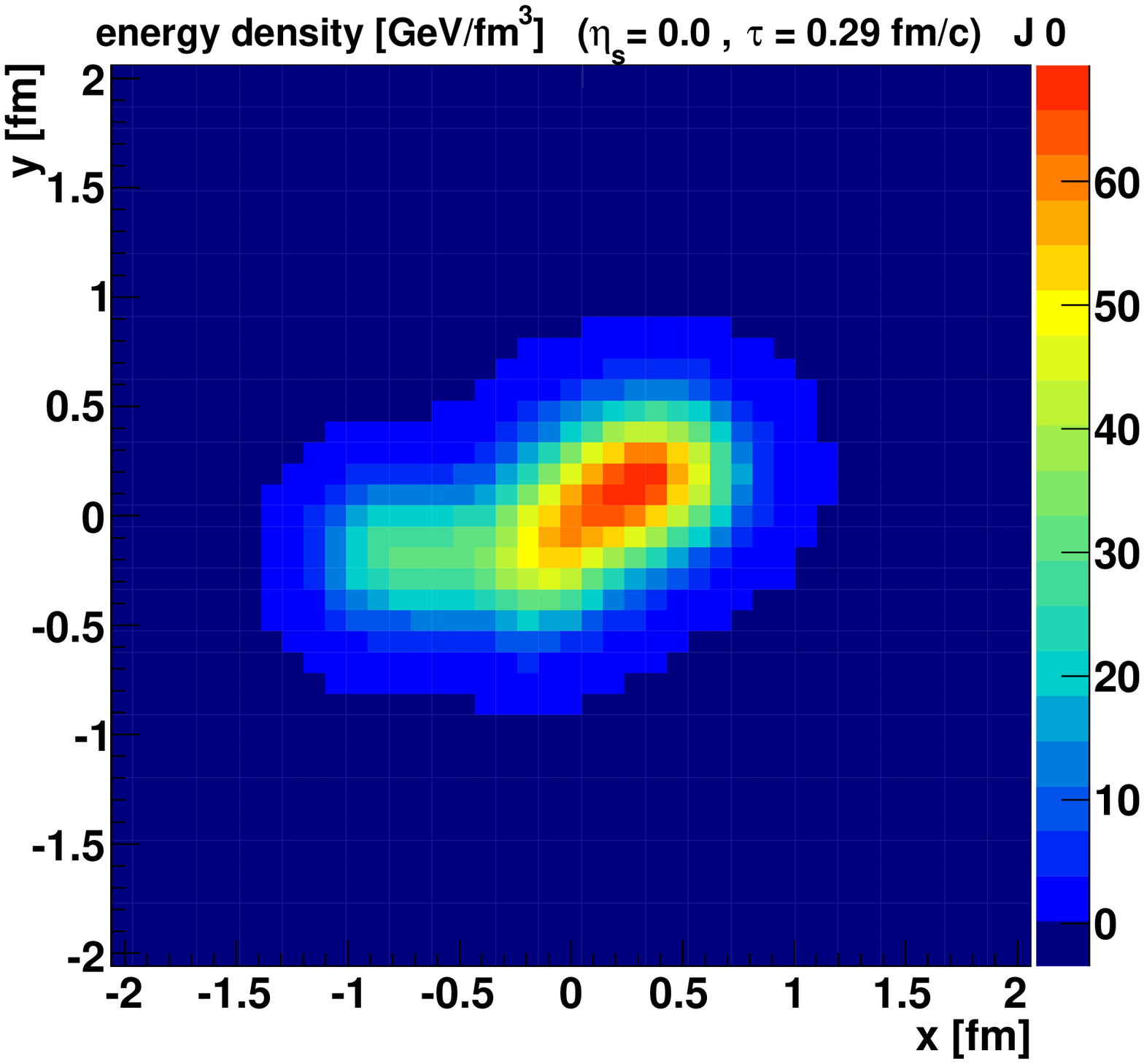}\includegraphics[scale=0.25]{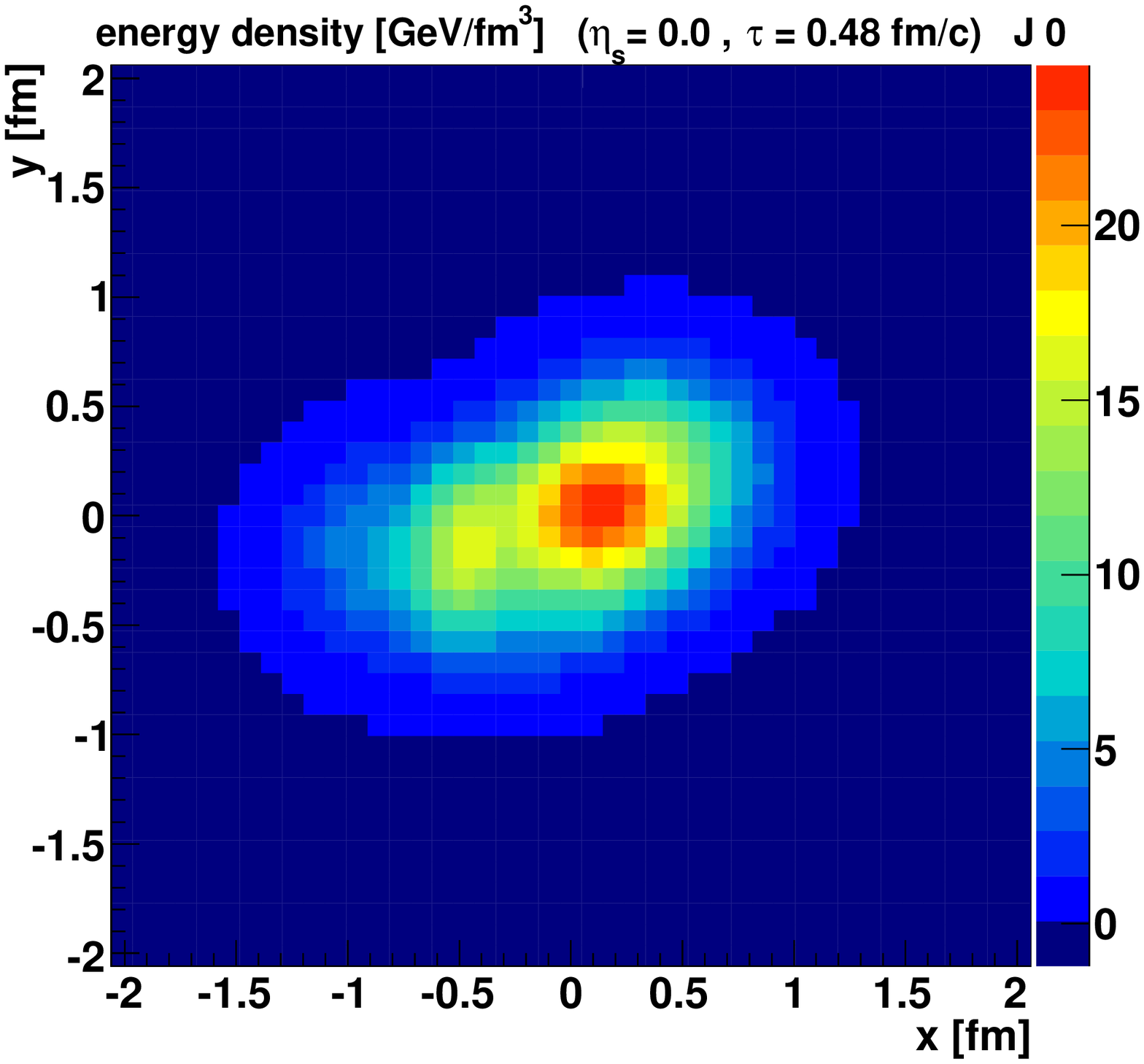}

\noindent \includegraphics[scale=0.25]{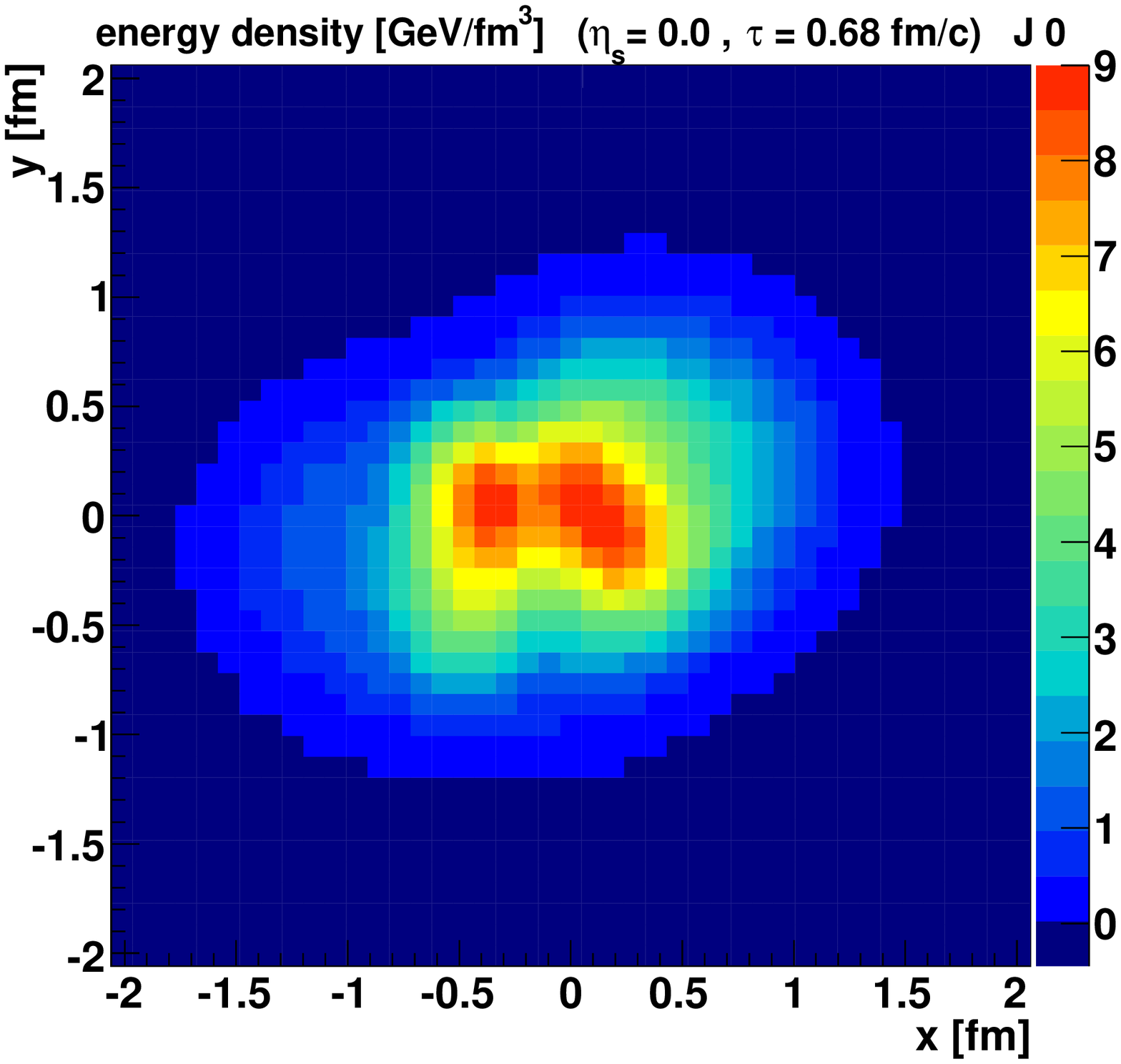}\includegraphics[scale=0.25]{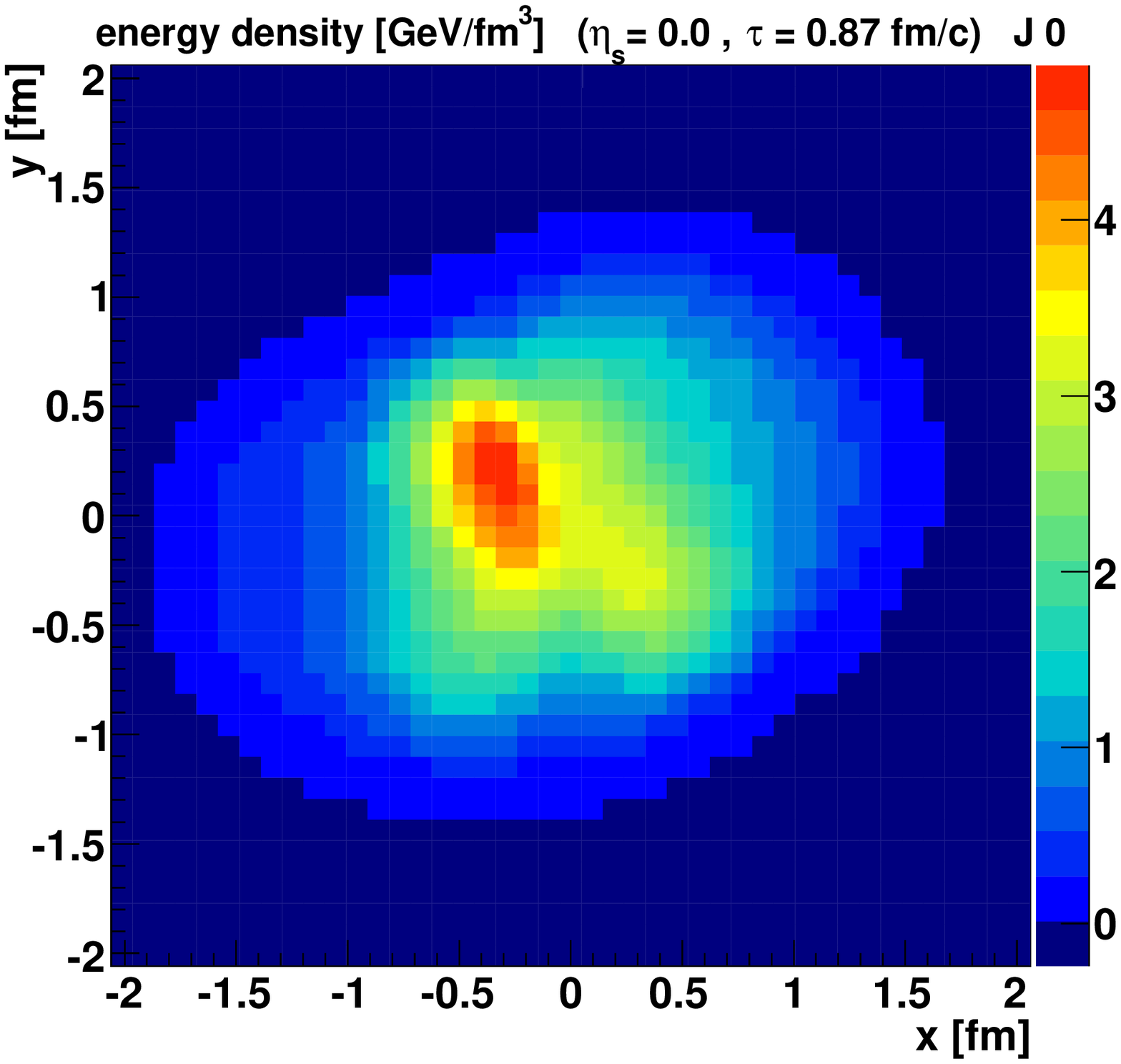}\includegraphics[scale=0.25]{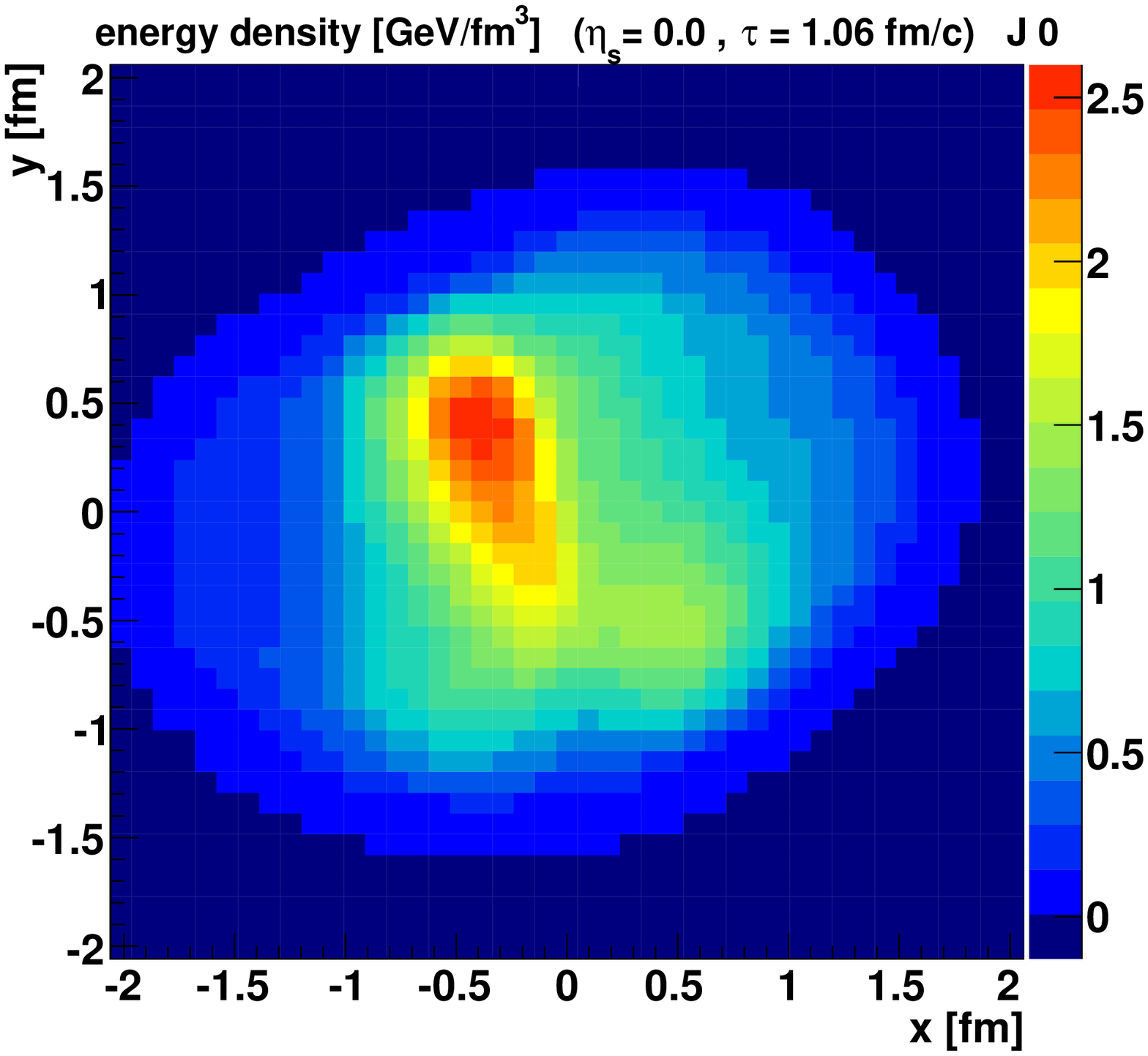}

\noindent \includegraphics[scale=0.25]{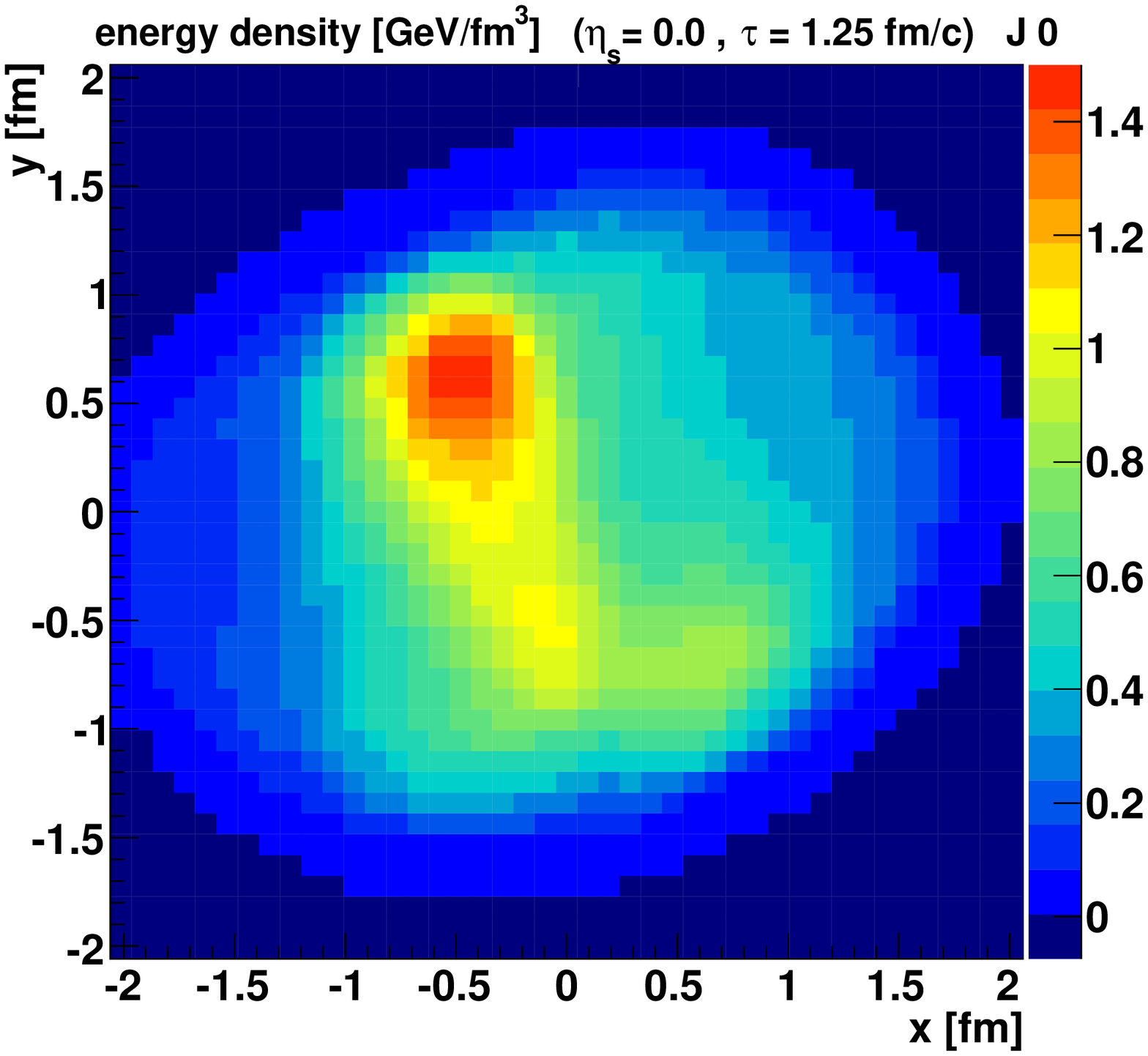}\includegraphics[scale=0.25]{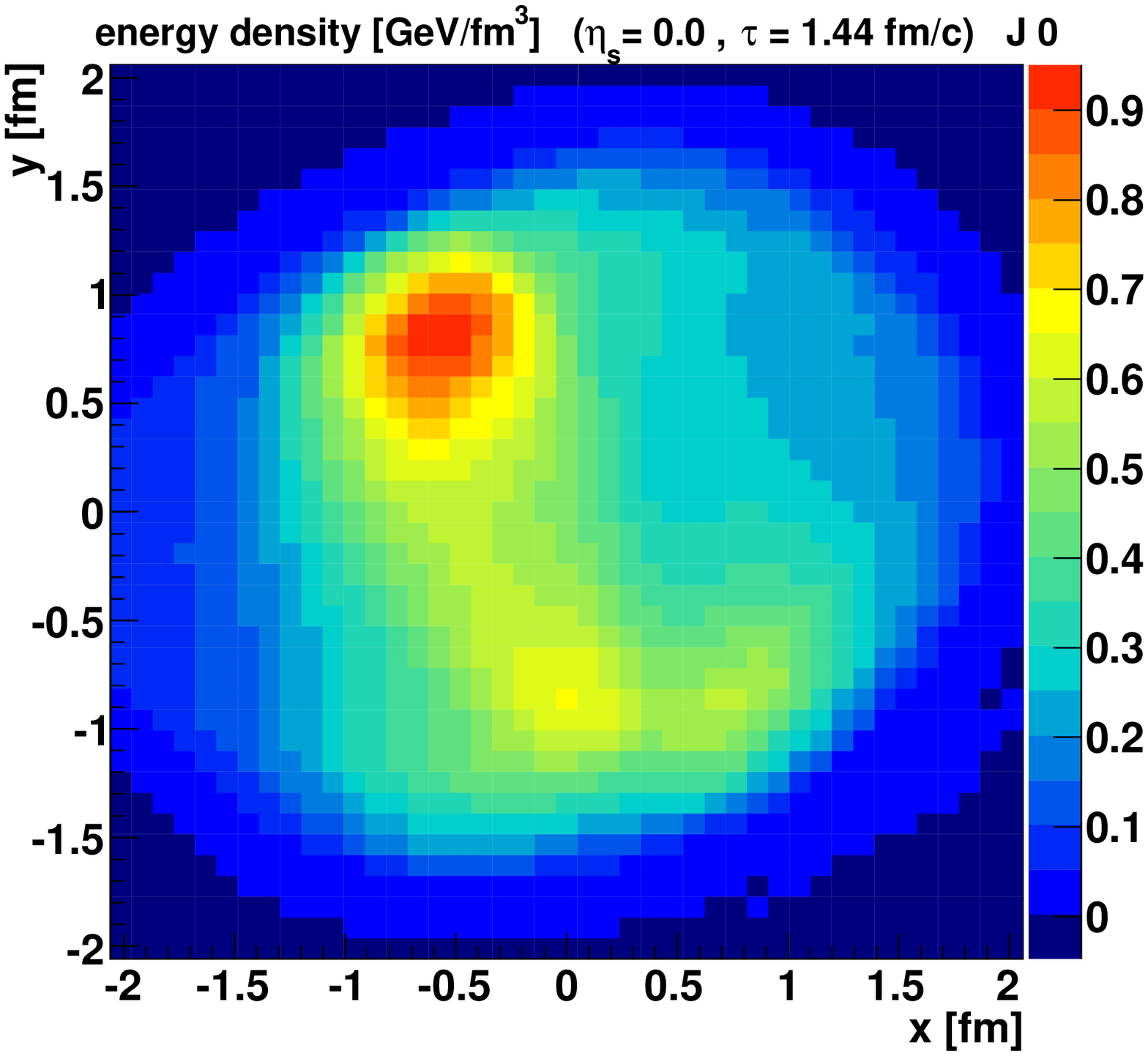}\includegraphics[scale=0.25]{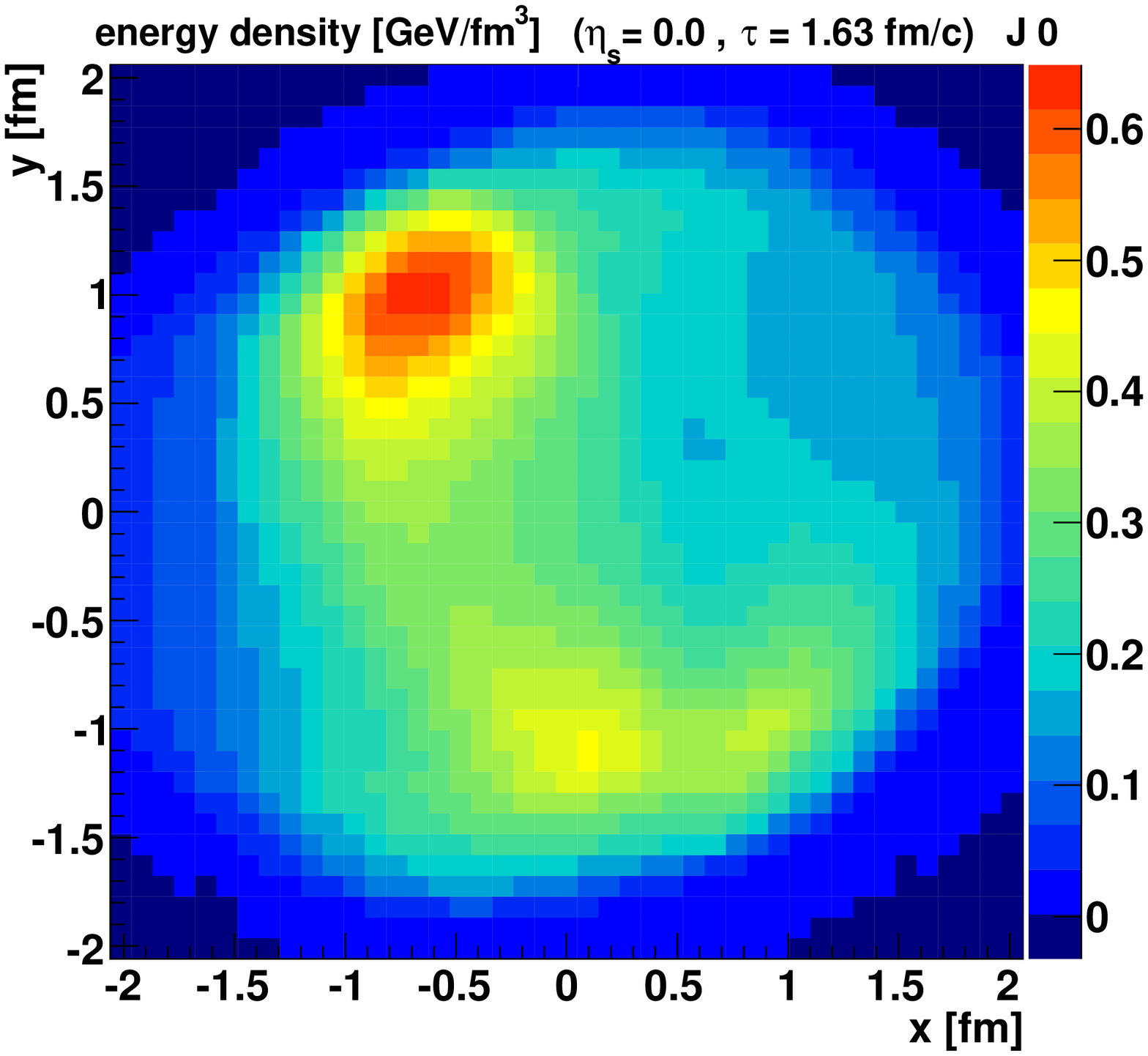}

\noindent \includegraphics[scale=0.25]{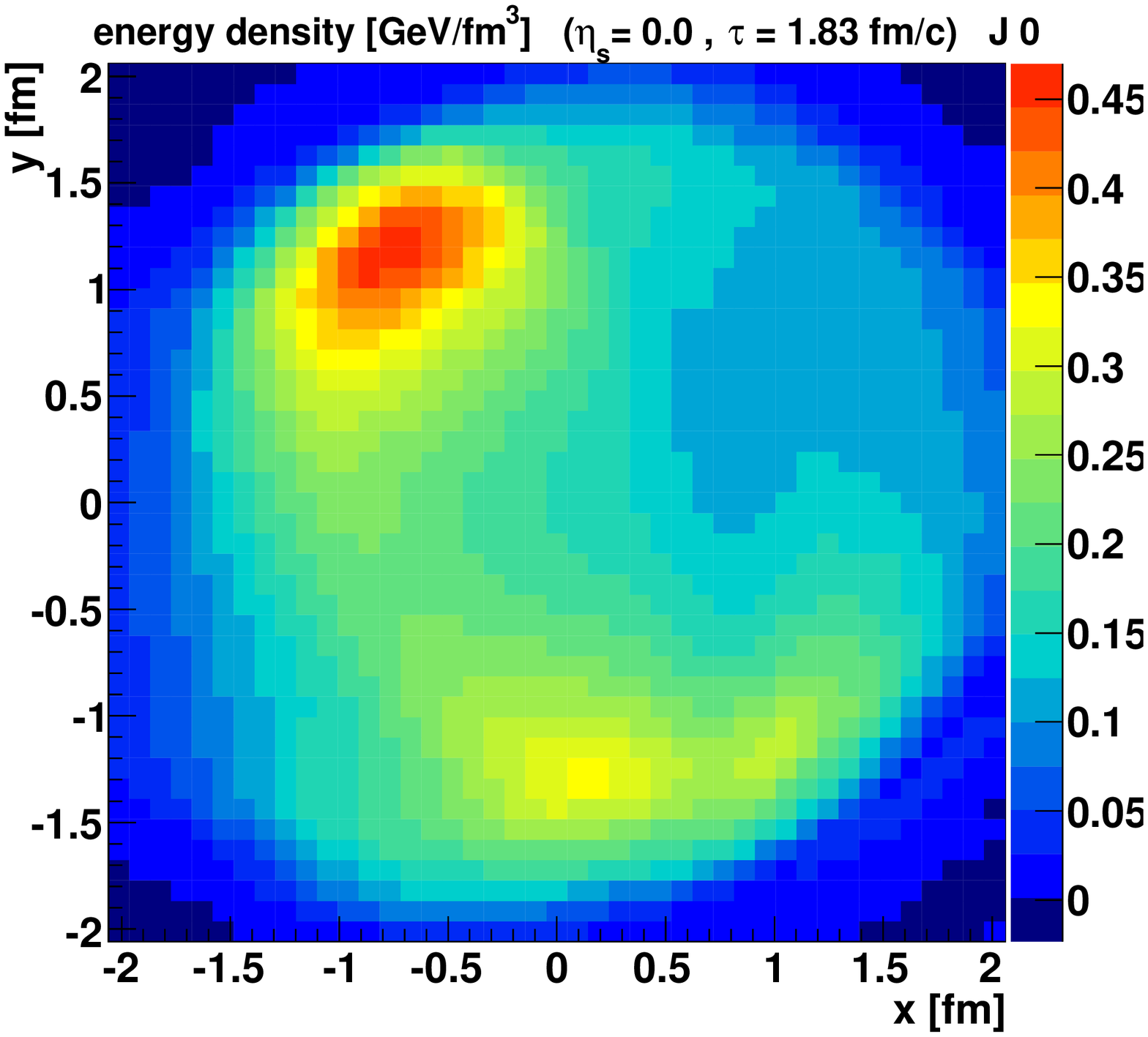}\includegraphics[scale=0.25]{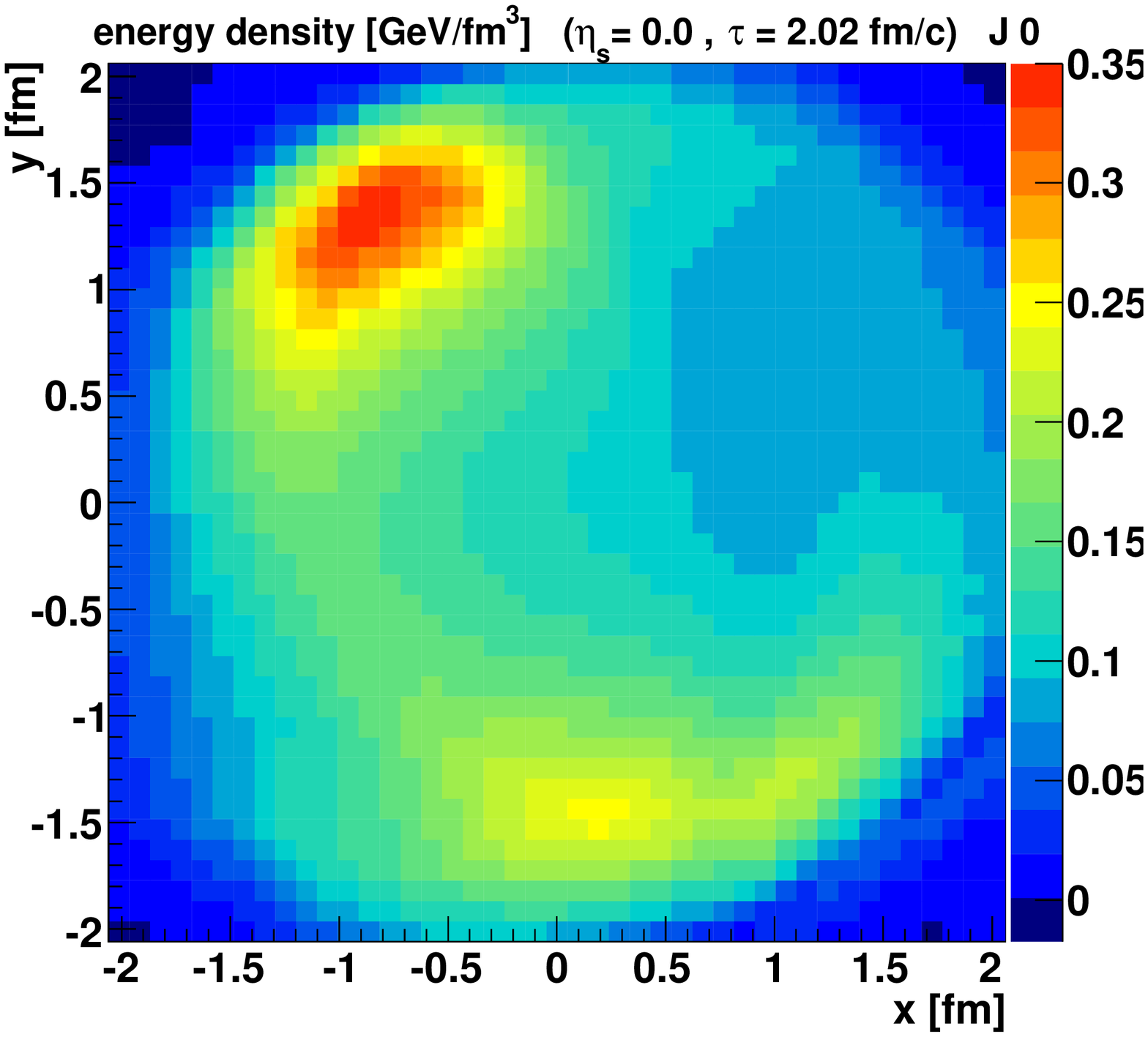}\includegraphics[scale=0.25]{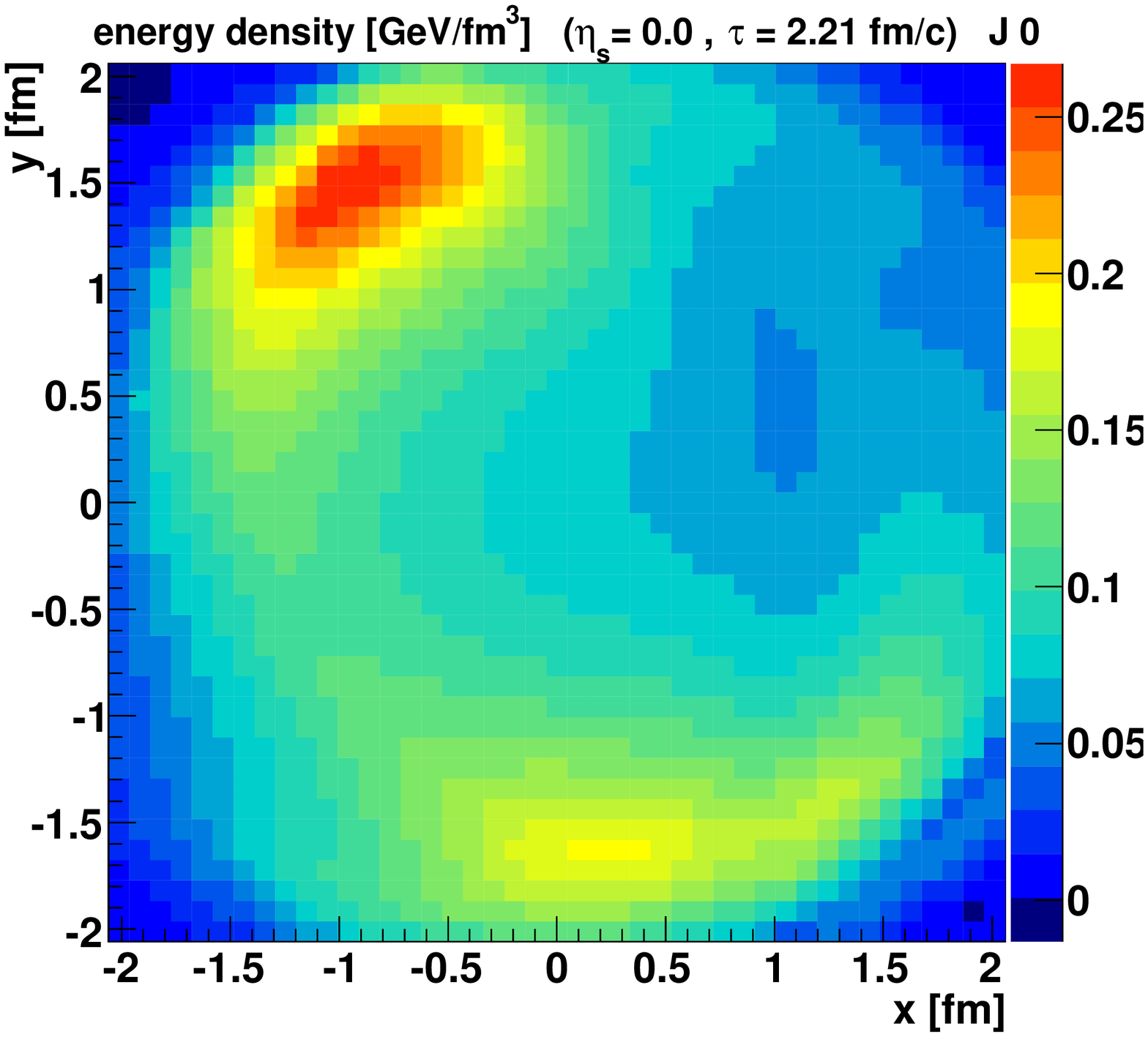}

\protect\caption{(Color online) Energy density distribution of the core in the transverse
plane ($x$, $y$ coordinates) for $z=0$ and for different proper
times. The upper left graph corresponds to the core as obtained from
the core-corona separation procedure.\label{fig:spacetime}}
\end{figure*}

EPOS3 \cite{epos3} is a universal model in the sense that for pp,
pA, and AA collisions, the same procedure applies, based on several
stages:
\begin{description}
\item [{\uline{Initial~conditions.}}] A Gribov-Regge multiple scattering
approach is employed (``Parton-Based Gribov-Regge Theory'' PBGRT
\cite{hajo}, see Fig. \ref{muscatt}), where the elementary object
(by definition called Pomeron) is a DGLAP parton ladder, using in
addition a CGC motivated saturation scale \cite{sat1} for each Pomeron,
of the form\textbf{ $Q_{s}\propto N_{\mathrm{part}}\,\hat{s}^{\lambda}$},
where $N_{\mathrm{part}}$ is the number of nucleons connected the
Pomeron in question, and $\hat{s}$ its energy. The parton ladders
are treated as classical relativistic (kinky) strings.
\item [{\uline{Core-corona~approach.}}] At some early proper time $\tau_{0}$,
one separates fluid (core) and escaping hadrons, including jet hadrons
(corona), based on the momenta and the density of string segments
(First described in \cite{core}, a more recent discussion in \cite{epos3}).
The corresponding energy-momentum tensor of the core part is transformed
into an equilibrium one, needed to start the hydrodynamical evolution,
see Fig. \ref{fig:spacetime}. This is based on the hypothesis that
equilibration happens rapidly and affects essentially the space components
of the energy-momentum tensor.
\item [{\uline{Viscous~hydrodynamic~expansion.}}] Starting from the
initial proper time $\tau_{0}$, the core part of the system evolves
according to the equations of relativistic viscous hydrodynamics \cite{epos3,yuri},
where we use presently $\eta/s=0.08$. A cross-over equation-of-state
is used, compatible with lattice QCD \cite{lattice,kw1}. 
\item [{\uline{Statistical~hadronization.}}] The ``core-matter''
hadronizes on some hypersurface defined by a constant temperature
$T_{H}$, where a so-called Cooper-Frye procedure is employed, using
equilibrium hadron distributions, see \cite{kw1}.
\item [{\uline{Final~state~hadronic~cascade.}}] After hadronization,
the hadron density is still big enough to allow hadron-hadron rescatterings.
For this purpose, we use the UrQMD model \cite{urqmd}.
\end{description}
The above procedure is employed for each event (event-by-event procedure).

Whereas our approach is described in detail in \cite{epos3}, referring
to older works \cite{hajo,core,kw1}, we confine ourselves here to
a couple of remarks, to selected items. The initial conditions are
generated in the Gribov-Regge multiple scattering framework. Our formalism
is referred to as ``Parton-Based Gribov-Regge Theory'' (PBGRT) and
described in very detail in \cite{hajo}, see also \cite{epos3} for
all the details of the present (EPOS3) implementation. The fundamental
assumption of the approach is the hypothesis that the S-matrix is
given as a product of elementary objects, referred to as Pomerons.
Once the Pomeron is specified (taken as a DGLAP parton ladder, including
a saturation scale), everything is completely determined. Employing
cutting rule techniques, one may express the total cross section in
terms of cut and uncut Pomerons, as sketched in fig. \ref{muscatt}.
The great advantage of this approach: doing partial summations, one
obtains expressions for partial cross sections $d\sigma_{\mathrm{exclusive}}$,
for particular multiple scattering configurations, based on which
the Monte Carlo generation of configurations can be done. No additional
approximations are needed. The above multiple scattering picture is
used for p-p, p-A, and A-A. 

Based on the PBGRT approach, we obtain in high multiplicity $pp$
collisions a large number of strings. The randomness of their transverse
positions leads to asymmetric energy density distributions in the
transverse plane at initial proper time $\tau_{0}$, as shown in the
left upper plot of Fig. \ref{fig:spacetime}, where we show the energy
density of the core for a particular (typical) event at space-time
rapidity $\eta_{s}=0$. In this example we observe an elongated (kind
of elliptical) shape. The other plots in Fig. \ref{fig:spacetime}
show the proper time evolution of the system. Mainly due to the strong
longitudinal expansion, the energy density values drop very fast.
Also a transverse expansion is visible, leading first to a more symmetric
shape, but then we see very clearly an expansion perpendicular to
the principal axis of the initial distribution, reflecting a typical
flow behavior, leading to non-zero harmonic flow coefficients and
ridges in dihadron correlation functions (see \cite{eposppridge}).

\noindent Detailed studies of pt spectra and azimuthal anisotropies
(dihadron corr., $v_{n}$) in pp and pA can be found in \cite{epos3,epos3b}.
All parameters have been fixed, there are no more free parameters
concerning the calculation discussed in this paper.

\section{Heavy quark production in EPOS3}

\begin{figure}
\noindent \begin{centering}
\includegraphics[scale=0.17]{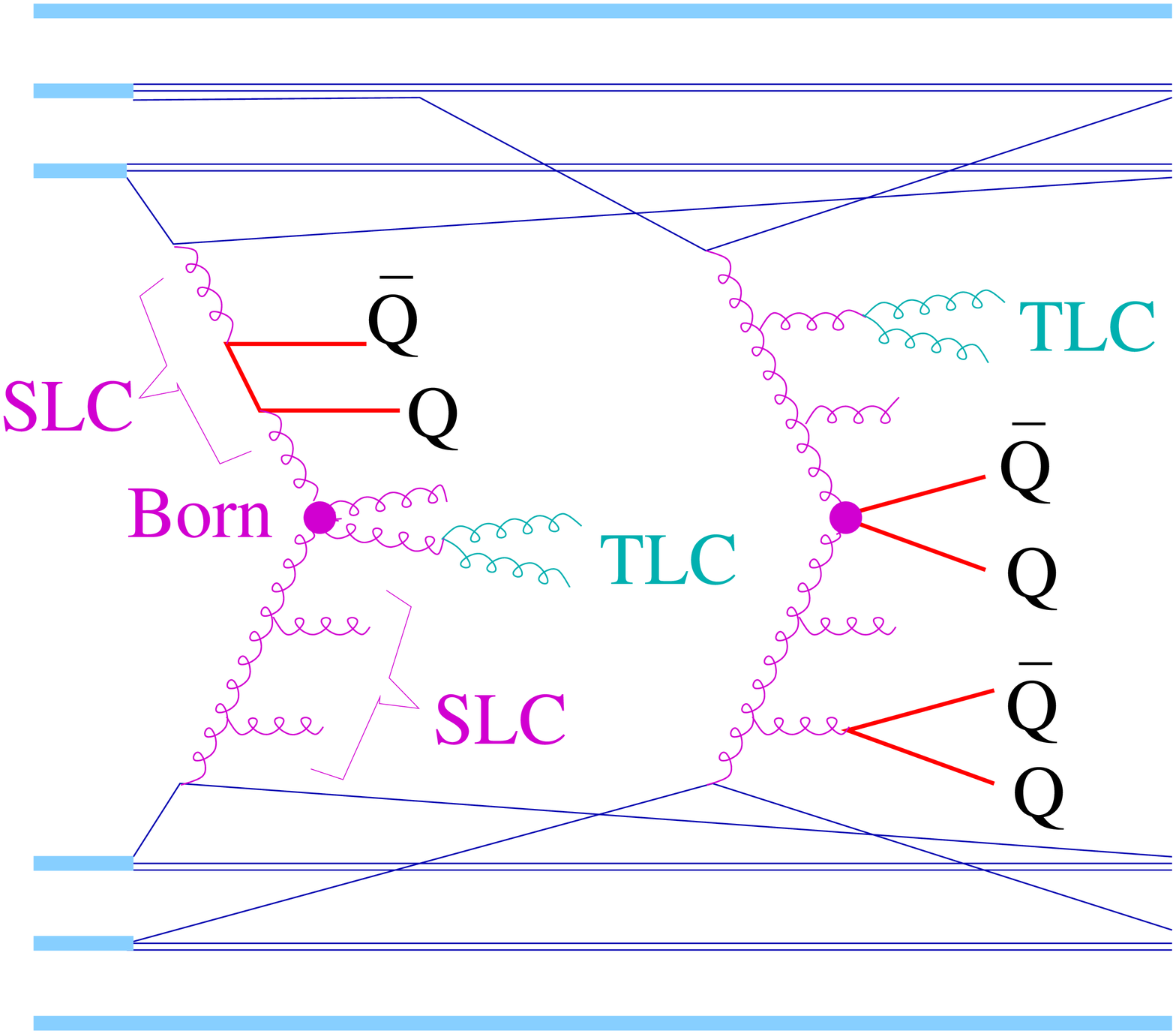}
\par\end{centering}

\protect\caption{(Color online) Heavy quark production in EPOS3. \label{fig:heavyquarks}}
\end{figure}

Heavy quarks (Q) are produced during the initial stage, in the PBGRT
formalism, in the same way as light quarks. We have several parton
ladders, each one composed of two space-like parton cascades (SLC)
and a Born process. The time-like partons emitted in the SLC or the
Born process are in general starting points of time like cascades
(TLC). In all these processes, whenever quark--antiquark production
is possible, heavy quarks may be produced. We take of course into
account the modified kinematics in case of non-zero quark masses (we
use $m_{c}=1.3,\,m_{b}=4.2$). In fig. \ref{fig:heavyquarks}, we
show several possibilities of heavy quark production in parton ladders.

In the present work, no interactions between these initially produced
heavy quarks and the fluid are considered. So all the D mesons are
originating from these initial hard processes.

$D$ meson production in the EPOS3 framework has been studied extensively,
comparing to data and other calculation, in ref. \cite{benjamin}.

\section{Correlating \textit{\large{}D} meson and charged particle multiplicities
in EPOS basic}

As discussed in the introduction, we try to understand the dependence
of the $D$ meson multiplicity on the charged particle multiplicity,
first for EPOS basic (without hydro). We study the case, where both
multiplicities refer to central rapidities ($|y|\leq0.5$ for the
$D$ mesons, and $|\eta|\leq1$ for the charged particles). As shown
in Tab. \ref{tab:definitions}, 
\begin{table}[b]
\begin{tabular}{|c|l|}
\hline 
$N_{\mathrm{ch}}$ & Charged particle multiplicity\tabularnewline
\hline 
$N_{D1}$ & D-meson multiplicity for $1<p_{t}\mathrm{[GeV/c}]<2\,$\tabularnewline
\hline 
$N_{D2}$ & D-meson multiplicity for $2<p_{t}\mathrm{[GeV/c}]<4$\tabularnewline
\hline 
$N_{D4}$ & D-meson multiplicity for $4<p_{t}\mathrm{[GeV/c}]<8\,$\tabularnewline
\hline 
$N_{D8}$ & D-meson multiplicity for $8<p_{t}\mathrm{[GeV/c}]<12$ \tabularnewline
\hline 
\end{tabular}

\protect\caption{Definitions of the variables $N_{\mathrm{ch}}$ and $N_{Di}$.\label{tab:definitions} }

\end{table}
we use the variables $N_{\mathrm{ch}}$ for the charged particle multiplicity,
and $N_{Di}$ for the $D$ meson multiplicities for different $p_{t}$
ranges. Experimental results exist as well for these particles individually,
but they seem to be identical, within the error bars. 

\begin{figure}
\noindent \begin{centering}
\includegraphics[scale=0.17]{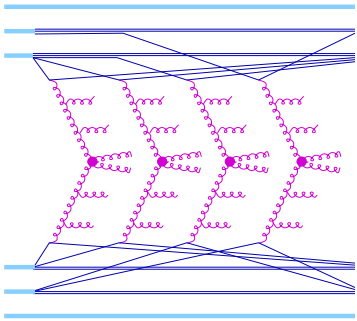}
\par\end{centering}

\protect\caption{(Color online) An example of a multiple scattering event in EPOS:
Four scatterings (parton ladders = cut Pomerons). \label{fig:multipleScattering}}
\end{figure}

Before discussing the results of actual calculations, let us make
some qualitative statements about what to expect from the basic element
of the EPOS approach, namely multiple scattering. We have in each
individual event a certain number of parton ladders (cut Pomerons),
as shown in fig. \ref{fig:multipleScattering}.
Each ladder contributes (roughly, on the average) the same to both
charged particle and charm production, so both corresponding multiplicities
are proportional to the number $N_{\mathrm{Pom}}$ of cut Pomerons:
\begin{equation}
N_{\boldsymbol{Di}}\propto N_{\boldsymbol{\mathrm{ch}}}\propto N_{\boldsymbol{\mathrm{Pom}}},\label{eq:linearity}
\end{equation}
which leads to a ''natural'' linear relation between the charged
particle multiplicity $N_{\mathrm{ch}}$ and the $D$ meson multiplicities
$N_{Di}$ (to first approximation). Although in reality, we will find
some deviation from these first approximation results, we will use
in the following nevertheless $N_{\mathrm{Pom}}$ as reference.

\begin{figure}
\noindent \begin{centering}
\includegraphics[angle=270,scale=0.24]{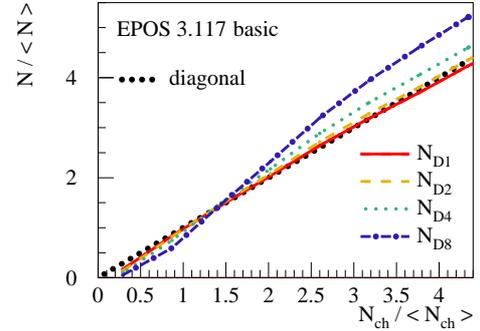}
\par\end{centering}

\protect\caption{(Color online) The calculation of $N_{Di}$ versus $N_{\mathrm{ch}}$
as obtained from EPOS basic. \label{fig:NDvsNch}}
\end{figure}

In fig. \ref{fig:NDvsNch}, 
we show the actual calculation of $N_{Di}$ versus $N_{\mathrm{ch}}$
(in EPOS basic). Indeed a roughly linear increase is observed, as
expected from eq. (\ref{eq:linearity}). Actually the increase is
even more than linear! (in particular for large $p_{t}$).

\begin{figure}
\noindent \begin{centering}
\includegraphics[angle=270,scale=0.24]{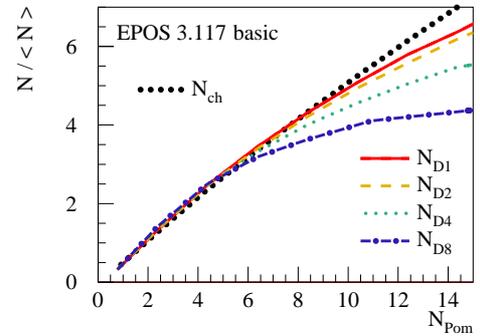}
\par\end{centering}

\protect\caption{(Color online) Average multiplicities (of charged particles and $D$
mesons) versus $N_{\mathrm{Pom}}$ as obtained from EPOS basic. \label{fig:NvsNpom}}
\end{figure}

A more than linear increase is nice (goes into the right direction
compared to data), but difficult to understand, in particular when
considering fig. \ref{fig:NvsNpom}, 

\noindent where we plot charged and $D$ meson multiplicities versus
the Pomeron number $N_{\mathrm{Pom}}$. Here, $D$ meson multiplicities
(in particular at high $p_{t}$) increase less than \emph{N}$_{\mathrm{ch}}$.

\begin{figure}[b]
\noindent \begin{centering}
\includegraphics[scale=0.28]{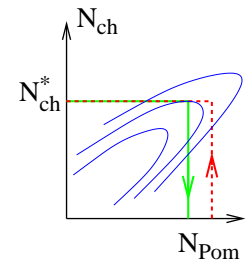}
\par\end{centering}

\protect\caption{(Color online) Contour plot of the number of events as a function
of $N_{\mathrm{ch}}$ and $N_{\mathrm{Pom}}$ (artists view). \label{fig:NEVvsNchNpom}}
\end{figure}

\noindent How to understand a more than linear $N_{D8}(N_{\mathrm{ch}})$
together with the fact that $N_{D8}$ increases much less with $N_{\mathrm{Pom}}$
than $N_{\mathrm{ch}}$? Crucial for this discussion are fluctuations.
There is certainly a strong correlation between $N_{\mathrm{ch}}$
and $N_{\mathrm{Pom}}$, but it is not a one-to-one correspondence,
as shown in a qualitative fashion in fig. \ref{fig:NEVvsNchNpom},
where we plot the number of events as a function of $N_{\mathrm{ch}}$
and $N_{\mathrm{Pom}}$. We see a ridge, which is narrow, but finite.
This results in the fact that a fixed Pomeron number $N_{\mathrm{Pom}}$,
which gives a particular $N_{\mathrm{ch}}$ as mean value, and the
average Pomerons number $N_{\mathrm{Pom}}^{*}$ for a fixed value
$N_{\mathrm{ch}}$ are not identical ($N_{\mathrm{Pom}}^{*}\neq$$N_{\mathrm{Pom}}$,
green and red arrows in the figure). Fig. (\ref{fig:NvsNpom}) is
therefore not really useful for discussing $N_{Di}(N_{\mathrm{ch}})$,
we have to look more into the details. 

We first define normalized multiplicities,
\begin{equation}
n=N/\left\langle N\right\rangle ,
\end{equation}
both for charged particles ($n_{\mathrm{ch}}$) and $D$ meson multiplicities
($n_{Di}$). In the following, we consider fixed values $n_{\mathrm{ch}}^{*}$
of normalized charged multiplicities. 

\begin{figure}
\noindent \begin{centering}
\includegraphics[angle=270,scale=0.24]{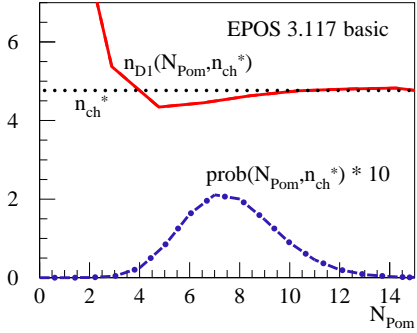}
\par\end{centering}

\protect\caption{(Color online) Pomeron number distribution at fixed charged multiplicity,
$\mathrm{prob}(N_{\mathrm{Pom}},n_{\mathrm{ch}}\!^{*})$ (blue line),
and number $n_{D1}(N_{\mathrm{Pom}},n_{\mathrm{ch}}\!^{*})$ of $D$
mesons (small $p_{t}$) for fixed $N_{\mathrm{Pom}}$ and $n_{\mathrm{ch}}\!^{*}$
as a function of the Pomeron number $N_{\mathrm{Pom}}$ (red line).
The dotted line represents the constant value $n_{\mathrm{ch}}\!^{*}$.
\label{fig:NchFixed}}
\end{figure}

\begin{figure}
\noindent \begin{centering}
\includegraphics[angle=270,scale=0.24]{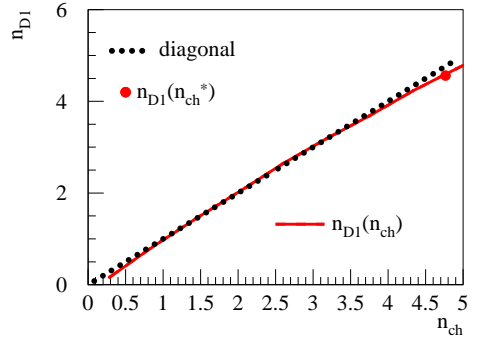}
\par\end{centering}

\protect\caption{(Color online) Average multiplicity $n_{D1}(n_{\mathrm{ch}})$ of
$D$ mesons (small $p_{t}$) as a function of $n_{\mathrm{ch}}$ (red
line) and the diagonal ($n_{D1}=n_{\mathrm{ch}}$, dotted line). The
red point refers to $n_{D1}(n_{\mathrm{ch}}^{*})$ for the particular
value of $n_{\mathrm{ch}}\!^{*}$ used in eq. (\ref{eq:NpomContributions}).
\label{fig:ND1vsNch}}
\end{figure}

Consider the average normalized $D$ meson multiplicity for the smallest
$p_{t}$ range, for some given $n_{\mathrm{ch}}^{*}$, referred to
as $n_{D1}(n_{\mathrm{ch}}^{*})$ , which may be written as
\begin{eqnarray}
n_{D1}(n_{\mathrm{ch}}^{*})=\sum_{N_{\mathrm{Pom}}}\mathrm{prob}(N_{\mathrm{Pom}},n_{\mathrm{ch}}\!^{*})\,\,\nonumber \\
\times\,n_{D1}(N_{\mathrm{Pom}},n_{\mathrm{ch}}\!^{*})\,\,\,\,,\label{eq:NpomContributions}
\end{eqnarray}
with $\mathrm{prob}(N_{\mathrm{Pom}},n_{\mathrm{ch}}\!^{*})$ being
the Pomeron number distribution at fixed $n_{\mathrm{ch}}^{*}$, and
$n_{D1}(N_{\mathrm{Pom}},n_{\mathrm{ch}}\!^{*})$ the number of $D$
mesons for fixed $N_{\mathrm{Pom}}$ and $n_{\mathrm{ch}}\!^{*}$.
The latter two curves (as obtained from EPOS basic) are plotted in
fig. \ref{fig:NchFixed} 
as a function of the Pomeron number $N_{\mathrm{Pom}}$ (red and blue
curve) together with the constant value of $n_{\mathrm{ch}}\!^{*}$,
indicated by the dotted line. In the range where the Pomeron distribution
is non-zero, the function $n_{D1}(N_{\mathrm{Pom}},n_{\mathrm{ch}}\!^{*})$
is roughly constant, and even close to the value $n_{\mathrm{ch}}\!^{*}$.
Using this ($n_{D1}(N_{\mathrm{Pom}},n_{\mathrm{ch}}\!^{*})\approx n_{\mathrm{ch}}\!^{*}$),
we get from eq. (\ref{eq:NpomContributions}):
\begin{equation}
n_{D1}(n_{\mathrm{ch}}^{*})\approx n_{\mathrm{ch}}\!^{*}.\label{eq:approxi}
\end{equation}
The result of the exact calculation is shown in fig. \ref{fig:ND1vsNch}
as red point. Also shown is the complete curve $n_{D1}(n_{\mathrm{ch}})$
as obtained from EPOS basic. Indeed, as indicated by eq. (\ref{eq:approxi}),
we get an almost perfect linear increase.

\begin{figure}[b]
\noindent \begin{centering}
\includegraphics[angle=270,scale=0.24]{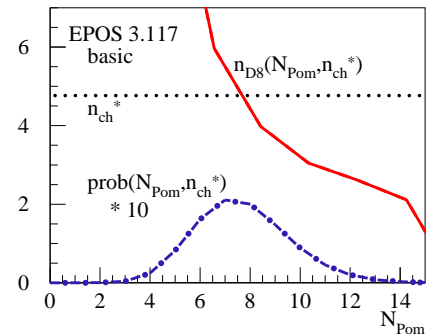}
\par\end{centering}

\protect\caption{(Color online) Pomeron number distribution at fixed charged multiplicity,
$\mathrm{prob}(N_{\mathrm{Pom}},n_{\mathrm{ch}}\!^{*})$ (blue line),
and number $n_{D8}(N_{\mathrm{Pom}},n_{\mathrm{ch}}\!^{*})$ of $D$
mesons (large $p_{t}$) for fixed $N_{\mathrm{Pom}}$ and $n_{\mathrm{ch}}\!^{*}$
as a function of the Pomeron number $N_{\mathrm{Pom}}$ (red line).
The dotted line represents the constant value $n_{\mathrm{ch}}\!^{*}$.
\label{fig:NchFixed-1}}
\end{figure}

\begin{figure}
\noindent \begin{centering}
\includegraphics[angle=270,scale=0.24]{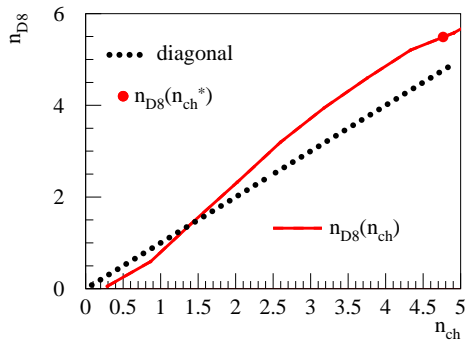}
\par\end{centering}

\protect\caption{(Color online) Average multiplicity $n_{D8}(n_{\mathrm{ch}})$ of
$D$ mesons (large $p_{t}$) as a function of $n_{\mathrm{ch}}$ (red
line) and the diagonal ($n_{D1}=n_{\mathrm{ch}}$, dotted line). The
red point refers to $n_{D8}(n_{\mathrm{ch}}^{*})$ for the particular
value of $n_{\mathrm{ch}}\!^{*}$ used in eq. (\ref{eq:NpomContributions-1}).
\label{fig:ND1vsNch-1}}
\end{figure}

Now we will study the average normalized $D$ meson multiplicity for
the largest $p_{t}$ range, for some given $n_{\mathrm{ch}}^{*}$,
which may be expressed as well in terms of the Pomeron number distribution
$\mathrm{prob}(N_{\mathrm{Pom}},n_{\mathrm{ch}}\!^{*})$ at fixed
$n_{\mathrm{ch}}^{*}$ and the number $n_{D8}(N_{\mathrm{Pom}},n_{\mathrm{ch}}\!^{*})$
of $D$ mesons for fixed $N_{\mathrm{Pom}}$ and $n_{\mathrm{ch}}\!^{*}$,
as 
\begin{eqnarray}
n_{D8}(n_{\mathrm{ch}}^{*})=\sum_{N_{\mathrm{Pom}}}\mathrm{prob}(N_{\mathrm{Pom}},n_{\mathrm{ch}}\!^{*})\,\,\nonumber \\
\times\,n_{D8}(N_{\mathrm{Pom}},n_{\mathrm{ch}}\!^{*})\,\,\,\,.\label{eq:NpomContributions-1}
\end{eqnarray}
The two curves representing $\mathrm{prob}(N_{\mathrm{Pom}},n_{\mathrm{ch}}\!^{*})$
and $n_{D8}(N_{\mathrm{Pom}},n_{\mathrm{ch}}\!^{*})$ are shown in
fig. \ref{fig:NchFixed-1}.
We see in the figure that $n_{D8}(N_{\mathrm{Pom}},n_{\mathrm{ch}}\!^{*})$
increases strongly towards small $N_{\mathrm{Pom}}$ with an inceasing
slope. Let us compare the expression of eq. (\ref{eq:NpomContributions-1})
with the corresponding sum (as a reference) where we use $n_{D8}(N_{\mathrm{Pom}},n_{\mathrm{ch}}\!^{*})=n_{\mathrm{ch}}\!^{*}$,
which would lead to $n_{D8}(n_{\mathrm{ch}}^{*})=n_{\mathrm{ch}}^{*}$.
For large $N_{\mathrm{Pom}}$, the contribution to the sum in eq.
(\ref{eq:NpomContributions-1}) will be less than the reference case,
but this is more than compensated at small $N_{\mathrm{Pom}}$. Therefore,
we have 
\begin{equation}
n_{D8}(n_{\mathrm{ch}}^{*})>n_{\mathrm{ch}}\!^{*},\label{eq:approxi-1}
\end{equation}
which is confirmed by the precise calculation shown in fig. \ref{fig:ND1vsNch-1}
as red point.
Also shown is the complete curve $n_{D8}(n_{\mathrm{ch}})$ as obtained
from EPOS basic. Indeed, as indicated by eq. (\ref{eq:approxi-1}),
we get a more than linear increase.

\begin{figure}
\noindent \begin{centering}
\includegraphics[angle=270,scale=0.24]{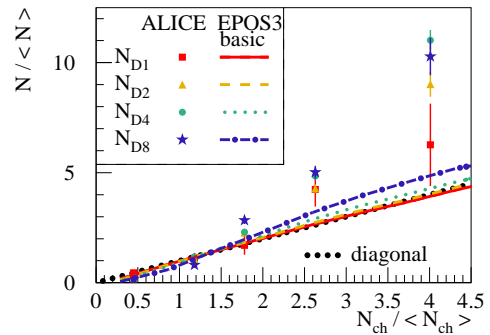}
\par\end{centering}

\protect\caption{(Color online) $D$ meson multiplicities versus the charged particle
multiplicity, both divided by the corresponding minimum bias mean
values. The different symbols and the notations $N_{D1}$, $N_{D2}$,
$N_{D4}$, $N_{D8}$ refer to different $p_{t}$ ranges: 1-2, 2-4,
4-8, 8-12 (in GeV), $N_{\mathrm{ch}}$ refers to the charged particle
multiplicity. We compare our calculation from EPOS basic (lines) to
ALICE data (points). \label{fig:NDvsNch_data-calc}}
\end{figure}

Let us summarize the main points of the preceding discussion: 
\begin{itemize}
\item The number of Pomerons fluctuates for given charged multiplicity.
\item The multiplicity \emph{N}$_{D8}$ of high transverse momentum $D$
mesons increases strongly towards small \emph{N}$_{\mathrm{Pom}}$
for given multiplicity, which simply means that it is favored to produce
high $p_{t}$ $D$ mesons for fewer (and more energetic) Pomerons.
\item This leads to a more than linear increase of the $D$ meson multiplicity
as a function of the charged particle one.
\item The effect is absent for low $p_{t}$ $D$ mesons, which show a linear
increase. 
\end{itemize}
The results of our calculation agree qualitatively with the trend
in the data, namely a more than linear increase, in particular for
high transverse momentum $D$ mesons. But the effect is actually too
small, as seen in fig. \ref{fig:NDvsNch_data-calc}, 
where we plot the $D$ meson multiplicities versus the charged particle
multiplicity, both for our calculation and data from ALICE \cite{alice1}.

\section{The influence of the hydrodynamical evolution}

\begin{figure}[b]
\noindent \begin{centering}
\includegraphics[angle=270,scale=0.24]{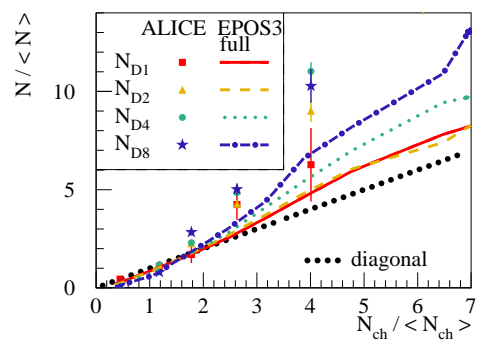}
\par\end{centering}

\protect\caption{(Color online) Same as fig. \ref{fig:NDvsNch_data-calc}, but here
we show the calulations based on the full EPOS model (with hydro).
\label{fig:NDvsNch_data-calc-1}}
\end{figure}

As seen in the previous chapter, we do see a more than linear increase
of $n_{Di}(n_{\mathrm{ch}})$\emph{ }in EPOS basic, but the effect
is too small. But anyhow, EPOS basic (w/o hydro) reproduces neither
spectra nor correlations, we have to consider the full approach, i.e.
EPOS with hydrodynamical evolution (with or without hadronic cascade
makes no difference). In fig. \ref{fig:NDvsNch_data-calc-1}, 
we plot again the $D$ meson multiplicities versus the charged particle,
EPOS3 compared to data, but here we refer to the calulations based
on the full EPOS model (with hydro). We see a significant non-linear
increase, much more pronounced as in the case of EPOS basic (without
hydro). 

How can we understand this increased non-linearity, due to the hydrodynamical
evolution? This is what we are going to discuss in the following. 

\begin{figure}
\noindent \begin{centering}
\hspace*{-0.3cm}\includegraphics[angle=270,scale=0.28]{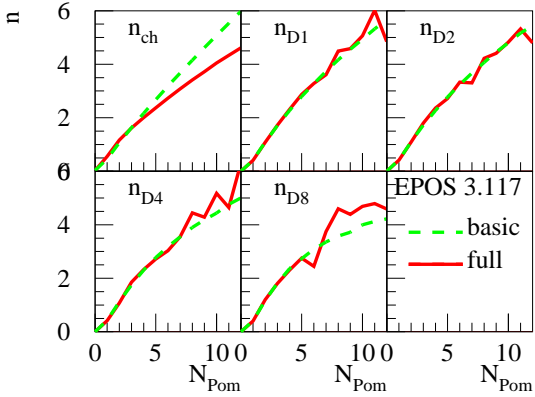}
\par\end{centering}

\protect\caption{(Color online) Average normalized multiplicities (of charged particles
and $D$ mesons) versus $N_{\mathrm{Pom}}$ obtained from EPOS basic
(dashed green lines) as compared to full EPOS (full red lines). \label{fig:NvsNpom-1} }
\end{figure}

\begin{figure}
\noindent \begin{centering}
\includegraphics[angle=270,scale=0.24]{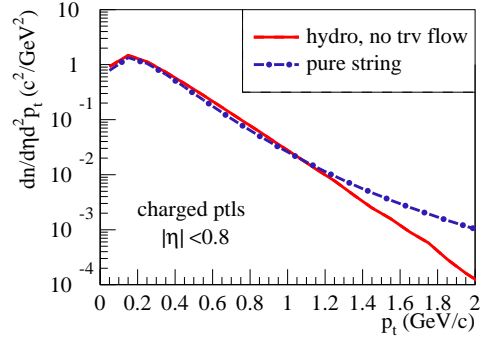}
\par\end{centering}

\protect\caption{(Color online) Transverse momentum spectra of charged particles in
a hydrodynamical scenario but with vanishing transverse flow (red
solid line), and from string fragmentation (blue dashed-dotted line).
\label{fig:twoScenarios}}
\end{figure}

In fig. \ref{fig:NvsNpom-1}, 
we plot the average normalized multiplicities $n_{\mathrm{ch}}$ and
$n_{Di}$ as a function of the Pomeron number $N_{\mathrm{Pom}}$.
We compare the results from EPOS basic and the full EPOS approach.
As expected, for the $D$ meson multiplicities the two calculations
are identical, due to the fact that the hydrodynamical evolution does
not affect $D$ meson production. The situation is quite different
concerning the charged particle multiplicities: Here, the multiplicities
from full EPOS (including hydro) are considerably below the results
from EPOS basic (without hydro). This makes a big effect, and leads
to a significant nonlinear increase of the $D$ meson multiplicities
versus the charged particle multiplicity. But important to keep in
mind:
\begin{itemize}
\item \uline{Not} the \uline{charm} production is \uline{increased}
with increasing ``collision activity'' (Pomeron number),
\item \noindent but the \uline{charged} \uline{particle} multiplicity
is \uline{reduced} when including a hydrodynamical expansion.
\end{itemize}
Why do we have such a multiplicity reduction due to the hydrodynamical
evolution? To understand this, we compare in fig.~\ref{fig:twoScenarios}
the transverse momentum spectra of charged particles for two different
scenarios: 
\begin{itemize}
\item Particle production in a hydrodynamical scenario, as used in full
EPOS, but with vanishing transverse flow, and 
\item Particle production from string fragmentation, as in EPOS basic. 
\end{itemize}
It can be seen that at small transverse momentum the two scenarios
give identical results (by accident). From this, we understand that
in a full hydrodynamical scenario with a strong transverse flow, part
of the available energy goes into flow rather than particle production,
reducing the multiplicity.

\section{Transverse momentum dependence }

Why is the non-linearity of $N_{Di}(N_{\mathrm{ch}})$ more pronounced
at high $p_{t}$ ? The naive expectation would be that the $N_{ch}$
reduction should affect all $D$ meson $p_{t}$ ranges in the same
way. But again we have to look more into the details. 

\begin{figure}
\noindent \begin{centering}
\hspace*{-0.7cm}\includegraphics[angle=270,scale=0.2]{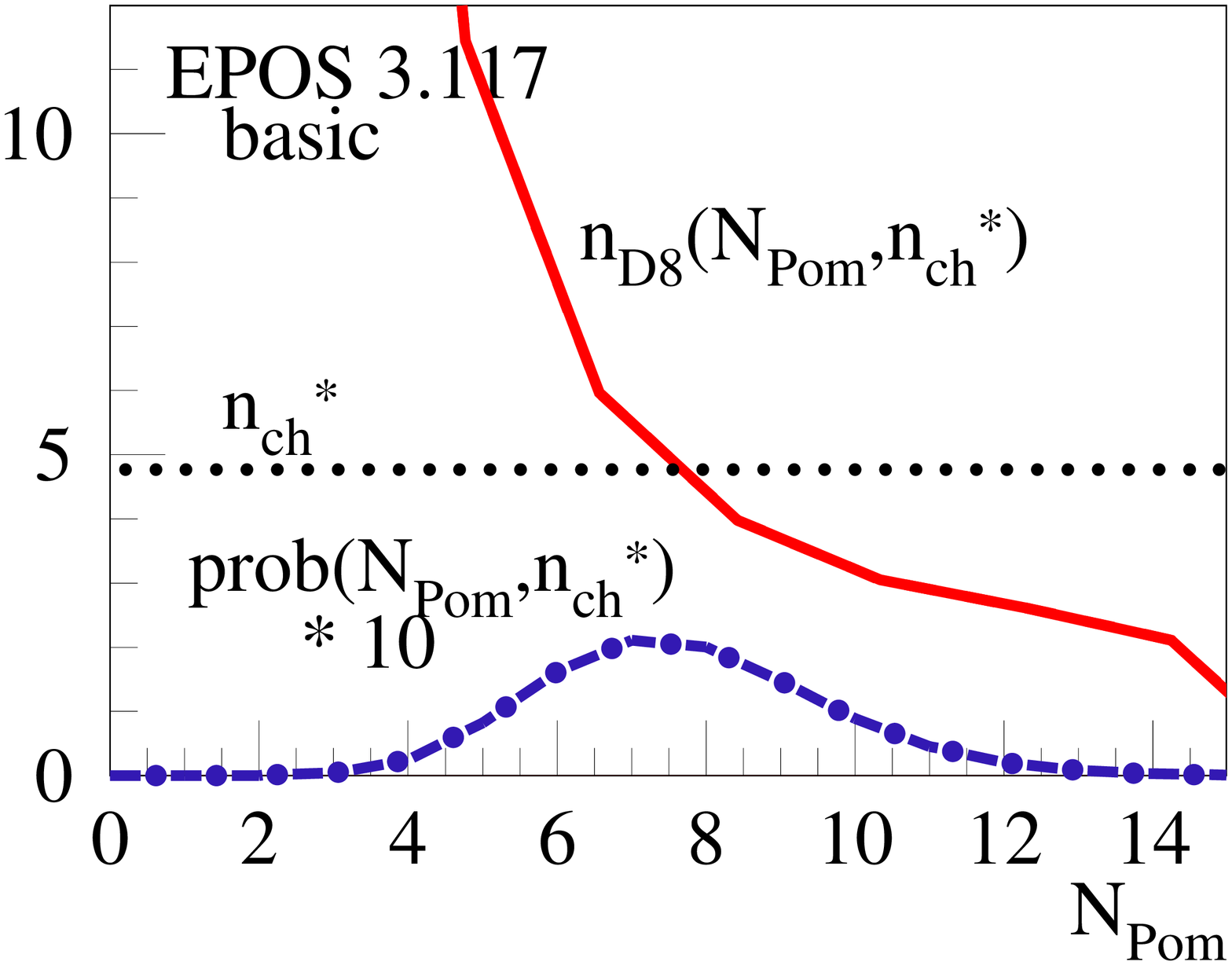}\hspace*{-0.45cm}\includegraphics[angle=270,scale=0.2]{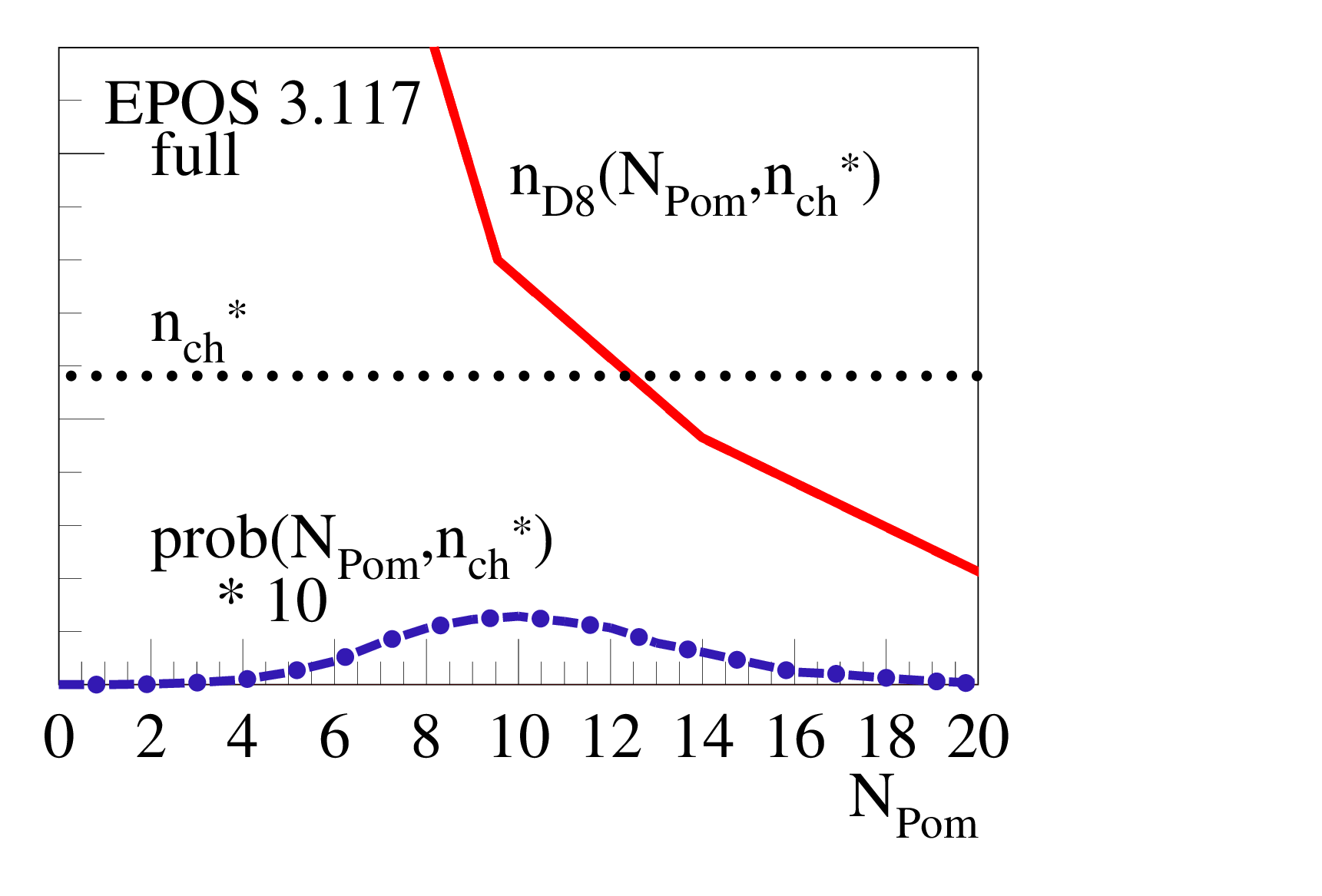}
\par\end{centering}

\protect\caption{(Color online) Pomeron number distribution at fixed charged multiplicity
(blue dashed-dotted lines), the number of $D$ mesons (large $p_{t}$)
at fixed charged multiplicity as a function of $N_{\mathrm{Pom}}$
(red full line), and the constant values of the charged multiplicity
(black dotted lines). We show results for EPOS basic (left plot) and
full EPOS (right plot) . \label{fig:NchFixed-1-1}}
\end{figure}

We remember the arguments related to fig.~\ref{fig:NchFixed-1},
where we plotted (for EPOS basic) the Pomeron number distribution
$\mathrm{prob}(N_{\mathrm{Pom}},n_{\mathrm{ch}}\!^{*})$ at fixed
charged multiplicity and the number $n_{D8}(N_{\mathrm{Pom}},n_{\mathrm{ch}}\!^{*})$
of $D$ mesons (large $p_{t}$) for given $N_{\mathrm{Pom}}$ and
$n_{\mathrm{ch}}\!^{*}$, as a function of the Pomeron number $N_{\mathrm{Pom}}$.
We could understand the non-linear increase of $n_{D8}(n_{\mathrm{ch}})$
to be due to fluctuations of the Pomeron numbers for fixed $n_{\mathrm{ch}}\!^{*}$
in connection with a strong increase of $n_{D8}(N_{\mathrm{Pom}},n_{\mathrm{ch}}\!^{*})$
towards small $N_{\mathrm{Pom}}$ ($D$ production is favored when
having fewer but more energetic Pomerons). We shown in fig. \ref{fig:NchFixed-1-1}
the same curves again (left plot) 
together with the corresponding curves for full EPOS (right plot).
We use the same fixed value of charged particle multiplicity for the
two scenarios ($N_{\mathrm{ch}\,(full)}^{*}=N_{\mathrm{ch}\,(basic)}^{*}$),
which leads to a somewhat bigger normalized multiplicity in case of
full EPOS ($n_{\mathrm{ch}\,(full)}^{*}>n_{\mathrm{ch}\,(basic)}^{*}$).
We see that including hydro leads to a shift towards bigger values
and a broadening of the Pomeron distribution. In addition, the curve
$n_{D8}(N_{\mathrm{Pom}},n_{\mathrm{ch}}^{*})$ crosses the constant
$n_{\mathrm{ch}}^{*}$ line even above the maximum of the Pomeron
number distribution. These two features enhance the effect (the non-linearity)
observed already for EPOS basic. For small transverse momenta this
``hydro-enhancement'' of the effect is absent (we do not have any
increase of $n_{D1}(N_{\mathrm{Pom}},n_{\mathrm{ch}}\!^{*})$ towards
small $N_{\mathrm{Pom}}$).

\section{Taking charged particle multiplicity at large rapidities}

We are still interested in the correlation of $D$ meson and charged
particle multiplicity, but here we consider the multiplicity at large
pseudo-rapidities, namely in the intervals $2.8<\eta<5.1$ and $\text{\textminus}3.7<\eta<\text{\textminus}1.7$,
which correspond to the Vzero detectors in ALICE. We refer to the
corresponding multiplicity as ``Vzero multiplicity'' using the symbol
$N_{vz}$, and $n_{vz}$ for the normalized multiplicity. The multiplicity
$n_{\mathrm{ch}}$ at central pseudo-rapidities, discussed in the
previous chapter, will be referred to as ``central multiplicity''.
\begin{figure}
\noindent \begin{centering}
\includegraphics[angle=270,scale=0.24]{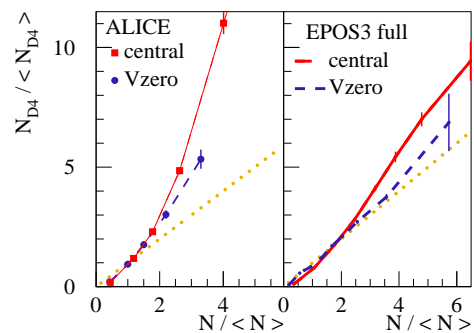}
\par\end{centering}

\protect\caption{(Color online) $D$ meson multiplicity $N_{D4}$ versus the Vzero
multiplicity (red solid lines) and versus the central multiplicity
(blue dashed lines). We show results from ALICE (left plot) and from
full EPOS simulations (right plot). \label{fig:NDvsNch_data-calc-2}}
\end{figure}
In fig. \ref{fig:NDvsNch_data-calc-2}, we plot the $D$ meson multiplicity
$N_{D4}$ versus the Vzero multiplicity (red solid line) and versus
the central muliplicity (blue dashed line). We show results from ALICE
(left plot) and from full EPOS simulations (right plot). Clearly visible
is the fact that (in both, data and simulation) the non-linear increase
is considerably reduced in case of Vzero multiplicity compared to
the central multiplicity. So it seems to matter whether the charged
particle multiplicity is consider in the same rapidity range as the
$D$ meson multiplicity or not. 

\begin{figure}[b]
\noindent \begin{centering}
\hspace*{-0.3cm}\includegraphics[angle=270,scale=0.24]{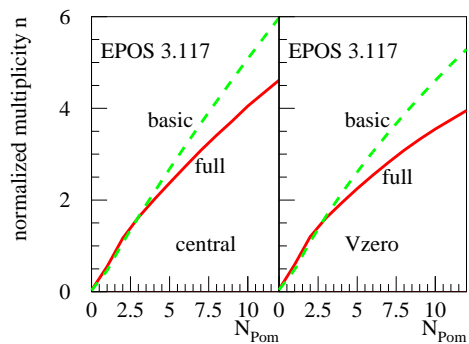}
\par\end{centering}

\protect\caption{(Color online) Average normalized charged particle muliplicity versus
$N_{\mathrm{Pom}}$ obtained from EPOS basic (dashed green lines)
as compared to full EPOS (full red lines). We show results for the
``central'' multiplicity (left) and the Vzero multiplicity (right).
\label{fig:NvsNpom-1-1} }
\end{figure}

How to understand this? Is there less ``hydro effect'' in case of
the Vzero multiplicity? The answer is ``no'', as can be seen in
fig. \ref{fig:NvsNpom-1-1}.
We see that also in the case of Vzero multiplicity, there is a strong
suppression of the multiplicity when considering hydrodynamical flow.
So the ``flow effect'' is the same for central and Vzero multiplicities.

\begin{figure}
\noindent \begin{centering}
\includegraphics[angle=270,scale=0.24]{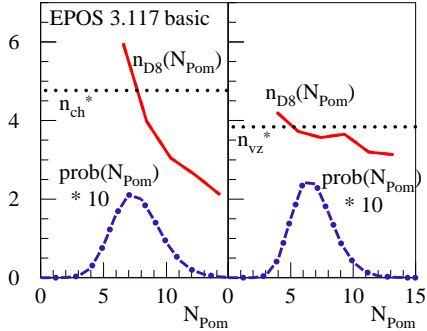}
\par\end{centering}

\protect\caption{(Color online) Pomeron number distribution at fixed charged multiplicity
(dashed-dotted blue lines) and number of $D$ mesons (large $p_{t}$)
for fixed $N_{\mathrm{Pom}}$ and fixed charged particle multiplicity
(solid red lines) as a function of the Pomeron number $N_{\mathrm{Pom}}$.
The dotted line represents the constant charged particle multiplicity.
We show results for the ``central multiplicity'' (left) and the
``Vzero multiplicity'' (right). \label{fig:NchFixed-1-2}}
\end{figure}

But the flow effect is only half of the story, it increases a primary
effect due to Pomeron number fluctuations. We therefore plot in fig.
\ref{fig:NchFixed-1-2}, 
the Pomeron number distribution at fixed charged multiplicity and
number of $D$ mesons (large $p_{t}$) for fixed $N_{\mathrm{Pom}}$
and fixed charged particle multiplicity, as a function of the Pomeron
number $N_{\mathrm{Pom}}$ for the two cases, namely ``central multiplicity''
(left) and the ``Vzero multiplicity'' (right). Whereas $n_{D8}$
increases strongly towards small $N_{\mathrm{Pom}}$ in the case of
``central multiplicity'', the corresponding curve for the ``Vzero
multiplicity'' (right plot) is almost flat, providing an essentially
linear increase of $n_{D8}(n_{\mathrm{vz}})$. The secondary effect
due to the hydro evolution will provide some non-linearity, but less
pronounced as compared to the case of central multiplicity, as seen
in fig. \ref{fig:NDvsNch_data-calc-2} (right plot, blue dashed line).
So the main difference between the case of ``central multiplicity''
(left) and the ``Vzero multiplicity'' is the fact that the latter
is uncorrelated with respect to the $D$ meson multiplicity.

\section{Summary }

We analyzed the dependence of $D$ meson multiplicities (in different
$p_{t}$ ranges) on the charged particle multiplicity in proton-proton
collisions at 7~TeV, using the EPOS3 approach. We find a non-linear
increase. Two issues play an important role: Multiplicity fluctuations
due to multiple scattering (realized via multiple Pomerons), and the
collective hydrodynamic expansion. Multiplicity fluctuations are important
since in particular high $p_{t}$ $D$ meson production at given (large)
charged particle multiplicity is very much enhanced for small Pomeron
numbers, which is responsible for the strong increase of the $D$
meson production with multiplicity. In addition, the effect is amplified
when turning on the hydrodynamical expansion, due to a reduction of
the charged particle multiplicity with respect to the model without
hydro. The non-linearity is reduced when considering multiplicities
at large pseudo-rapidities, not competing with $D$ meson production.
Our results are very robust with respect to many details of the modeling
(and not just specific to EPOS3), they essentially depend on very
basic features of the reaction mechanism in proton-proton collisions,
namely having multiple scattering and a multiplicity reduction due
to collective effects. The data discussed in this paper contain therefore
valuable information.
\begin{acknowledgments}
This research was carried out within the scope of the GDRE (European
Research Group) ``Heavy ions at ultrarelativistic energies''. B.G.
acknowledges the financial support by the TOGETHER project of the
Region of ``Pays de la Loire''. B. G. gratefully acknowledges generous
support from Chilean FONDECYT grants 3160493.\end{acknowledgments}

\end{document}